\documentclass[journal,10pt]{IEEEtran}
\usepackage{subfigure}
\PassOptionsToPackage{subfigure}{tocloft}
\usepackage{CJK}
\usepackage{algorithm}
\usepackage{algorithmic}
\usepackage{amsmath}
\usepackage{amssymb}
\usepackage{psfrag}
\usepackage{graphicx}
\usepackage{epstopdf}
\usepackage{multirow}
\usepackage{stfloats}
\usepackage{cite}
\usepackage{color}
\usepackage[normalem]{ulem}
\usepackage{arydshln}
\usepackage{enumerate}
\usepackage{bbm}
\usepackage{bm}
\title{Weighted Sum-Rate Maximization for Reconfigurable Intelligent Surface Aided Wireless Networks}
\author{
Huayan Guo, \IEEEmembership{Member, IEEE},  Ying-Chang Liang, \IEEEmembership{Fellow, IEEE},\\  Jie Chen, \IEEEmembership{Student Member, IEEE}, and Erik G. Larsson, \IEEEmembership{Fellow, IEEE}
\thanks{This work was supported by the National Natural Science Foundation of China under Grant U1801261, 61631005, and 61571100.  (\textit{Corresponding author: Ying-Chang Liang.})}
\thanks{H. Guo is with Shenzhen Research Institute, Hong Kong University of Science and Technology (HKUST), Shenzhen 518000, China, and he is also with the Department of Electronic and Computer Engineering, HKUST, Hong Kong 999077, China (e-mail: guohuayan@pku.edu.cn).}
\thanks{Y.-C. Liang and J. Chen are with the Center for Intelligent Networking and Communications (CINC), University of Electronic Science and Technology of China (UESTC), Chengdu 611731, China (e-mails:  liangyc@ieee.org; chenjie.ay@gmail.com).}
\thanks{E. G. Larsson is with the Department of Electrical Engineering (ISY), Link${\ddot{{\rm o}}}$ping University, SE-581 83 Link${\ddot{{\rm o}}}$ping, Sweden (email: erik.g.larsson@liu.se).}
}

\begin{document}

\maketitle
%\newpage

\begin{abstract}
Reconfigurable intelligent surfaces (RIS) is a promising solution to build a programmable wireless environment via steering the incident signal in fully customizable ways with reconfigurable passive elements. In this paper, we consider a RIS-aided multiuser multiple-input single-output (MISO) downlink communication system. Our objective is to maximize the weighted sum-rate (WSR) of all users by joint designing the beamforming at the access point (AP) and the phase vector of the RIS elements, while both the perfect channel state information (CSI) setup and the imperfect CSI setup are investigated. For perfect CSI setup, a low-complexity algorithm is proposed to obtain the stationary solution for the joint design problem by utilizing the fractional programming technique. Then, we resort to the stochastic successive convex approximation technique and extend the proposed algorithm to the scenario wherein the CSI is imperfect. The validity of the proposed methods is confirmed by numerical results. In particular, the proposed algorithm performs quite well when the channel uncertainty is smaller than 10\%.
\end{abstract}


\section{Introduction}
\emph{Reconfigurable intelligent surface} (RIS), also known as intelligent reflection surface, is an artificial  structure consisting of passive radio elements, each of which could adjust the reflection of the incident electromagnetic waves with unnatural properties \cite{survey2019,WuQQmagzine,Debbah2019IRSMISO,Liang2019JCIN,Tan2018SRA,Liu2019metasurface,cuiTJ2017metasurface}.
Moreover, owing to the passive structure,  the power consumption is extremely low, and there is nearly no additional thermal noise added during reflecting.
As a result, the RIS attracts more and more attentions in academia and industry with vast application prospect, e.g., wireless power transfer \cite{Mishra2019CEpowerT,wuqq2019SWIPT}, physical layer security \cite{zhangruiwcl,Chenjie2019access,xuweiSecrercyRate}, cognitive radio network \cite{TanCR}, etc.
%optical communications \cite{najafi2019optical}, etc.
Among them, one of the most promising applications is to improve the quality-of-service of users in the wireless communication system suffering from unfavorable propagation conditions  \cite{Liaskos2018magzineIRS,Bjornson2019massiveMIMO,Renzo2019position,Hum2014ReflectarrayReview,Renzo2019RIScomparetoRelay}.
%Besides, the RIS has also shown potential to enhance other specific applications,

\begin{figure}
[!t]
\centering
\includegraphics[width=.9\columnwidth]{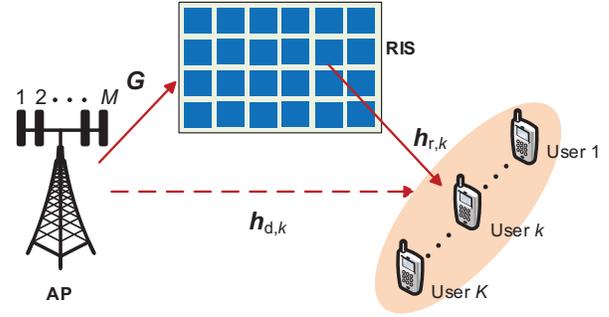}
\caption{The RIS-aided  multiuser MISO communication system.}
\label{IRS_system}
%\vspace{-1.5em}
\end{figure}

In this paper, we investigate the RIS-aided \emph{multiple-input single-output} (MISO) multiuser downlink communication system as shown in {\figurename~\ref{IRS_system}}, in which a multi-antenna \emph{access point} (AP) serves multiple single-antenna mobile users.
The direct links between the AP and the mobile users may suffer from deep fading and shadowing, and the RIS improves the propagation conditions by providing high-quality virtual links from the AP to the users.
While RIS resembles  a full-duplex {amplify-and-forward}  relay \cite{Larsson2014FDrelay,IRSvsDFrelay}, it forwards the RF signals via passive reflection, and thus has advantages in both energy- and cost- efficiency.
The objective of this paper is to maximize the \emph{weighted sum-rate} (WSR) of the mobile users by jointly optimizing the beamforming at the AP and the phase coefficients of the RIS elements.

\subsection{Related Works}
The system  in this paper has already investigated by some  early-attempt works, in which different objectives  are considered while
most works assume that the perfect \emph{channel state information} (CSI) of all  involved channels is available.
In \cite{zhangrui2018GcomIRS} and \cite{zhangruiIRS}, the  transmit power of the AP is minimized by decomposing the joint optimization problem into two subproblems: one is the conventional power-minimization problem in MIMO system, and the other is for the RIS phase vector optimization. Then the phase optimization problem is solved via \emph{semidefinite relaxation} (SDR) technique.
Although this alternating optimization approach achieves quite good performance, the main shortcoming is that the proposed algorithm cannot obtain the stationary solution, and the complexity is a little high especially for large-size RIS.
In \cite{Yuen2018ZF} and \cite{YuenChauIRS}, the energy efficiency is maximized, while employing \emph{zero-forcing} (ZF) precoding at the AP.
Since the ZF precoding completely cancels the inter-user interference, the power allocation at the AP and the phase optimization at RIS can be well decoupled.
However, the ZF precoding may as well amplify the  background noise, and the performance may be severely compromised when the channel is ill-conditioned.
Unfortunately, the derivations in \cite{YuenChauIRS} are not applicable directly for other precoding schemes.

Another important issue for the RIS-aided system is the channel estimation.
It is known from \cite{zhangrui2018GcomIRS,zhangruiIRS,Yuen2018ZF,YuenChauIRS} that, to optimize the phase vector of the RIS, the system needs high-accuracy CSI about the AP-RIS channel and the RIS-user channels, respectively.
However, to obtain perfect CSI is not always possible, since the RIS is passive without channel sensing capability in typical setup.
This challenge has been addressed by \cite{Jinshi2019TVTsurfaceAverage} and \cite{Dennah2019IRS} via exploiting statistical CSI.
Specifically, the single user system is investigated in \cite{Jinshi2019TVTsurfaceAverage}, and the average received SNR is maximized while assuming that the \emph{line-of-sight} (LoS) component of the channel is known.
In \cite{Dennah2019IRS}, multiuser system is considered in which all the users are located in the same cluster whose spatial correlation relation is known by the system.
Then, the max-min fairness problem is investigated by means of large dimensional random matrix theory.
However, the performance of these methods in \cite{Jinshi2019TVTsurfaceAverage} and \cite{Dennah2019IRS} depends heavily  on the channel model assumptions, as well as the objective functions investigated.

\subsection{Contributions}
In this paper, we first assume that perfect CSI is available to study the ultimate performance of the RIS-aided system.
The formulated problem looks mathematically similar to the WSR maximization problem in the hybrid digital/analog precoding \cite{Letaief2016manifoldHybrid,Ayach2014sparseprecoding,YuWei2016HybridBeam} in massive MIMO systems.
Nevertheless, the main difference is that, the RIS can only control and optimize the behavior of the wireless environment, and has no capability to suppress inter-user interference.
Due to that, the beamforming design at the AP and phase optimization at the RIS are deeply coupled, and the convergence speed of the alternating optimization approach is slow.
Therefore, the computational complexity in each iteration step should be low and scalable to the number of RIS elements.
To tackle this issue, we decompose the original problem into four disjoint blocks by utilizing the  \emph{fractional programming} (FP) technique \cite{YuWei2018FP2}. Subsequently, low-complexity algorithm is designed based on the non-convex \emph{block coordinate descent} (BCD) method \cite{nonconvexBCD2013LuoZQ}.

Then, we address the imperfect CSI issue.
Specifically, we assume that the AP may perfectly know the combined channel for beamforming design, since this knowledge can be obtained via conventional channel estimation method and protocol, and the antenna number of the AP may be not huge in the femtocell network \cite{Liaskos2018magzineIRS}. %Therefore, the beamforming may be designed by conventional method with perfect CSI given RIS phase vector.
However, the system only has partial knowledge about the channels related to the RIS phase optimization.
Hence, we design the phase of the RIS to maximize the average WSR for the incoming channel realizations.
This problem formulation is more general than those in \cite{Jinshi2019TVTsurfaceAverage} and \cite{Dennah2019IRS}, since it is independent to the structure of channel model assumptions.
Besides, although similar problem formulation can be found in the hybrid  precoding problem \cite{AnLiu2014HybridBeam,AnLiu2016HybridcodeBeam,AnLiu2018SSCPHybrid,AnLiu2019TTHybridmulticell}, the coupled optimization variables here make this problem much more complicated to be solved.
%The phase optimization problem here is a two-timescale  stochastic optimization problem, and similar problem formulation can also be found in the hybrid  precoding problem \cite{AnLiu2014HybridBeam,AnLiu2016HybridcodeBeam,AnLiu2018SSCPHybrid,AnLiu2019TTHybridmulticell}.
Fortunately,  we show that the proposed algorithm for perfect CSI cases can be extended to the imperfect CSI setup,
by utilizing the recently proposed stochastic  \emph{successive convex approximation} (SCA) technique \cite{ZQLuo2016SSUM,LiuATSP2018onlineSSCA}.

The main contributions of this work are summarized as follows:
\begin{itemize}
\item Firstly, this paper is one of the early attempts to study the WSR maximization problem for the RIS-aided multiuser downlink MISO system, and both the perfect and imperfect CSI setups are investigated.
\item Secondly, for perfect CSI setup, a BCD based method is proposed to carry out the stationary solution for the joint beamforming design and RIS phase optimization problem. The complexity of the proposed algorithm is much lower than the conventional approach.
\item Finally, the proposed algorithm is extended to the imperfect CSI cases. Numerical results verify that the proposed algorithm may perform well, when the channel  uncertainty is smaller than $10\%$.\footnote{{\textcolor{blue}{The source code of this paper will be uploaded online, when the paper is published.}}}
\end{itemize}

\subsection{Organization and Notations}
The rest of the paper is organized as follows. Section \ref{system model} outlines the system model and formulates the joint optimization problem.
In Section \ref{Sec:AO_ideal}, conventional alternating optimization approach is presented for the joint optimization problem under perfect CSI setup.
Then, in Section \ref{Sec:BCD_ideal}, a low-complexity algorithm is proposed based on the non-convex BCD technique.
Next, in Section \ref{Sec:Imperfect}, the proposed algorithm is extended to the imperfect CSI setup.
Simulation results are provided in Section \ref{simulation} to verify the effectiveness of the proposed algorithms, and Section \ref{conclusion} concludes the paper.

The notations used in this paper are listed as follows. ${\mathbb E}[\cdot]$ denotes statistical expectation.
${\cal{CN}}(\mu, \sigma^2)$ denotes the \emph{circularly symmetric complex Gaussian} (CSCG) distribution with mean $\mu$ and variance $\sigma^2$.
${\bf{I}}_{M}$ denotes the $M \times M$ identity matrix.
For any general matrix ${\bf G}$, $g_{i,j}$ is the $i$-th row and $j$-th column element.
${\bf G}^\ast$, ${\bf G}^{\rm T}$ and ${\bf G}^{\rm H}$ denote the conjugate, the transpose and the conjugate transpose of ${\bf G}$, respectively.
For any vector ${\bf w}$ (all vectors in this paper are column vectors), $w_i$ is the $i$-th element, and $\|{\bf w}\|$ and $|{\bf w}\|_{\rm F}$ denotes the Euclidean norm and the Frobenius norm, respectively.
%The quantity $\max(x,y)$ and $\min(x,y)$ denote the maximum and minimum between two real numbers $x$ and $y$, respectively.
$|x|$ denotes the absolute value of a complex number $x$, and ${\rm{Re}}\{x\}$ is its real part.

\section{System Model and Problem Formulation}\label{system model}
\subsection{Channel Model}
This paper investigates a RIS-aided multiuser MISO communication system, using modeling that   substantially follows \cite{zhangruiIRS}.
As shown in {\figurename~\ref{IRS_system}}, the system consists of one AP equipped with $M$ antennas, one RIS which has $N$ reflection elements, and $K$ single-antenna users.
We assume that all the channels experience quasi-static flat-fading.
The baseband equivalent channels from AP to user $k$, from AP to RIS, and from RIS to user $k$ are denoted by ${\bf h}_{{\rm d},k} \in {\mathbb C}^{M\times 1}$, ${\bf G} \in {\mathbb C}^{N\times M}$, and ${\bf h}_{{\rm r},k} \in {\mathbb C}^{N \times 1}$, respectively.
The phase-shift matrix  is defined as a diagonal matrix ${\bf \Theta}={\rm diag}(\theta_1, \cdots, \theta_n, \cdots, \theta_N)$,
where $\theta_n= e^{\jmath \varphi_n}$ is the phase of the $n$-th reflection element on RIS.
The reflection operation on the $n$-th RIS element resembles multiplying the incident signals with $\theta_n$, and then forwarding these composite signals as if from a point source.

Denote the transmit data symbol to user $k$ by $s_k$, which is independent random variables with zero mean and unit variance.
Then, the transmitted signal at the AP can be expressed as
\begin{equation*}
{\bf x}=\sum_{k=1}^K {\bf w}_k s_k,
\end{equation*}
where ${\bf w}_k \in {\mathbb C}^{M\times 1}$ is the corresponding transmit beamforming vector.

The signal received at user $k$ is expressed as
\begin{equation*}
\begin{aligned}[b]
{y}_k&= \underbrace{ {\bf h}_{{\rm d},k}^{\rm H} {\bf x} }_{\text {Direct~link}}
+\underbrace{ {\bf h}_{{\rm r},k}^{\rm H} {\bf \Theta} {\bf G}  {\bf x} }_{{\text {RIS-}}{\text {aided}}~{\text {link}}}
+u_k\\
&=\left({\bf h}_{{\rm d},k}^{\rm H}+{\bf h}_{{\rm r},k}^{\rm H} {\bf \Theta} {\bf G}  \right) \sum_{k=1}^K {\bf w}_k s_k +u_k
,
\end{aligned}
\end{equation*}
where $u_k \sim {\cal{CN}}(0,\sigma_0^2) $ denotes the \emph{additive white Gaussian noise} (AWGN) at the $k$-th user receiver.
To make above expression more tractable, we further define ${\bm \theta}= [\theta_1, \cdots, \theta_N]^{\rm H}$ and ${\bf H}_{{\rm r},k}={\rm diag}({\bf h}_{{\rm r},k}^{\rm H}){\bf G} \in {\mathbb C}^{N\times M}$, and then the received signal ${y}_k$ is equivalently represented as
\begin{equation}\label{equ:downlink_signal2}
{y}_k=  \left({\bf h}_{{\rm d},k}^{\rm H}+{\bm \theta}^{\rm H} {\bf H}_{{\rm r},k}\right) \sum_{k=1}^K {\bf w}_k s_k +u_k
.
\end{equation}

The $k$-th user  treats all the signals from other users (i.e., $s_1,\cdots,s_{k-1},s_{k+1},\cdots,s_K$) as interference.
Hence, the decoding SINR of $s_k$ at user $k$ is
\begin{equation}\label{equ:downlink_SINR}
{\gamma}_k= \frac{\left|({\bf h}_{{\rm d},k}^{\rm H}+{\bm \theta}^{\rm H} {\bf H}_{{\rm r},k}){\bf w}_k  \right|^2 }
{\sum_{i=1, i \neq k}^K \left|({\bf h}_{{\rm d},k}^{\rm H}+{\bm \theta}^{\rm H} {\bf H}_{{\rm r},k}){\bf w}_i  \right|^2+\sigma_0^2}
.
\end{equation}
In addition, the transmit power constraint of AP is
\begin{equation}\label{equ:power_constrant_BS}
\sum_{k=1}^K \|{\bf w}_k\|^2 \leq P_{\rm T}
.
\end{equation}

\subsection{Discussion on Channel Estimation}\label{system model_CSI}
Generally, to optimize the phase vector $\bm \theta$, the RIS-aided system should estimate ${\bf h}_{{\rm d},k}$, $\bf G$, and ${\bf h}_{{\rm r},k}$, respectively.
However, to obtain the high-accurate CSI is a key challenging problem, since the dimension of ${\bf h}_{{\rm d},k}$ and $\bf G$ grows linearly with $N$.
In existing literature, there are three main methods to estimate the CSI:
\begin{itemize}
\item  \emph{Brute-Force Method}: In \cite{Debbah2019IRSMISO}, a brute-force method is proposed, in which the CSI with respect to each RIS element is estimated sequentially by the AP while turning off other elements. The main drawback of this method is that the training overhead is huge (proportional to $N$). Then, in \cite{Zhangrui2019IRSOFDM}, this method is modified by grouping the adjacent elements to reduce the training overhead.
\item  \emph{Compressive-Sensing Method}: In \cite{YuanXJ2019LISChannelestimation}, a compressive-sensing based method is proposed by exploiting the low-rank property of the RIS-aided link to further reduce the training overhead.
\item  \emph{Semi-passive RIS}: In \cite{Taha2019IRSChannelE}, a semi-passive structure  is suggested by integrating active elements on the RIS which have the channel estimation capability. Then, it is shown that the training overhead may become negligible by exploiting deep learning  and compressive sensing tools.
\end{itemize}

%
%How to obtain CSI at IRS is a difficult task.
%Some early-attempts can be found in \cite{Taha2019IRSChannelE} and \cite{Liaskos2019IRSestimation}, in which a channel construct approach is proposed to obtain the full CSI with low training overhead based on compressive sensing tools.

\subsection{Problem Formulation}
In this paper, our objective is to maximize the WSR of all the users by jointly designing the transmit beamforming  at the AP and the phase vector   at RIS, subject to the transmit power constraint in \eqref{equ:power_constrant_BS}.
In addition, two setups are investigated for different assumptions on the CSI.

\subsubsection{Perfect CSI}
We first consider an ideal setup in which the  CSI of all channels involved is perfectly known.
The algorithms proposed for this setup may serve as a benchmark to study the ultimate performance of the system, as well as providing training labels for the machine learning based designs \cite{HuangDLRIS2019,Liaskos2019NeuralNetIRS}.

Let ${\bf W}=[{\bf w}_1,{\bf w}_2, \cdots, {\bf w}_K] \in {\mathbb C}^{M\times K}$.
The WSR maximization problem is formulated as
\begin{subequations}
\begin{align}
{\mathcal{P}}{\text{(A)}}\quad \max_{{\bf W}, {\bm \theta}} \quad &  f_{\rm A}({\bf W}, {\bm \theta})=\sum_{k=1}^K \omega_k \log (1+{\gamma}_k) \notag \\
{\bf s.t.} \quad
& |\theta_n| =1, \quad \forall n=1,\cdots,N, \label{equ:P1c1}\\
& \sum_{k=1}^K \|{\bf w}_k\|^2 \leq P_{\rm T}, \label{equ:P1c2}
%& \sum_k^K p_k \leq P_{\rm T}. \label{equ:P1c3}
\end{align}
\end{subequations}
where the weight $\omega_k$ is used to represent the priority of user $k$.
Since the optimal solution is irrelevant to the base of the logarithm function, we use the natural logarithm throughout the paper. %for ease of notation.

Despite the conciseness of ${\mathcal{P}}{\text{(A)}}$, the joint beamforming and phase optimization problem is generally much more difficult than the power minimization problem in \cite{zhangruiIRS}, and the ZF transmission based design in \cite{YuenChauIRS}, since the optimization variables ${\bf W}$ and  ${\bm \theta}$ are deeply coupled in the non-convex objective function.
In addition, as $N$ is usually large in practice, we prefer an algorithm with lower complexity which is scalable to $N$, while the complexity of the SDR technique adopted by \cite{zhangruiIRS} is ${\mathcal O}(N^6)$  which is a little high.
%Also note that, although similar joint optimization problem has been investigated by \cite{zhangruiIRS} for the power minimization objective and solved by the SDR technique, we prefer an algorithm with lower complexity which is scalable for $N$, as $N$ is usually large in practice.

\subsubsection{Imperfect CSI}
{\textcolor{black}{
We consider the TDD based transmission frame structure for the RIS-aided communication system as illustrated in {\figurename~\ref{IRS_frame}}. Specially, one time slot for RIS configuration is inserted between two traditional TDD transmission frames. After $\bm \theta$ is configured, the rest system design is totally the same as the traditional communication system.
}

\begin{figure}
[!ht]
\centering
\includegraphics[width=1.0\columnwidth]{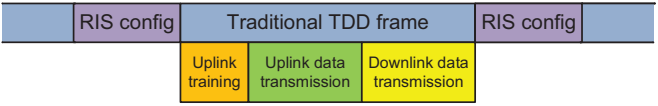}
\caption{The TDD based transmission frame structure for the RIS-aided communication system.}
\label{IRS_frame}
%\vspace{-1.5em}
\end{figure}

As such, to design $\bf W$, the AP only requires the knowledge of the combined channel information as follows:
\begin{equation}
{\bf h}_k={\bf h}_{{\rm d},k}+{\bf H}_{{\rm r},k}^{\rm H} {\bm \theta},
\end{equation}
whose dimension is $M$ (irrelevant to $N$). We still assume that ${\bf h}_k$ for all users are perfectly known. This is reasonable as the antenna number $M$ at the AP in a  femtocell network is usually not large, e.g., the AP in the smart-home application scenario is usually equipped with 2 to 8 transmit antennas.

{\textcolor{black}{
Therefore, in this setup, the key task is optimizing $\bm \theta$.
However, in order to configure $\bm \theta$, the channel coefficients ${\bf h}_{{\rm d},k}$, $\bf G$, and ${\bf h}_{{\rm r},k}$ should be estimated separately.
According to the references in Section \ref{system model_CSI}, these channel coefficients are difficult to be estimated in high accuracy with short training pilots.
In addition, the pilots and data symbols for channel estimation all come from the previous uplink-transmission slots, which also introduces estimation error.
Hence, denote the imperfect channel estimations by ${{\hat{\bf h}}_{{\rm d},k}}$, ${\hat{\bf G}}$ and ${{\hat{\bf h}}_{{\rm r},k}}$, respectively.
The  estimation error could be expressed by \cite{Cheng2006TWCimperfectCSI,Taesang2006ITimperfectCSI,Dabbagh2008TcomimperfectCSI}:
\begin{align*}
{\bf z}_{{\rm d},k}&={{{\bf h}}_{{\rm d},k}}-{{\hat{\bf h}}_{{\rm d},k}},\\
{\bf Z}_{{\rm G}}&={\bf G}-{\hat{\bf G}}, \\
{\bf z}_{{\rm r},k}&={{{\bf h}}_{{\rm r},k}}-{{\hat{\bf h}}_{{\rm r},k}}.
\end{align*}
If minimum \emph{mean square error} (MSE) estimation is applied, the error ${\bf z}_{{\rm d},k}$, ${\bf Z}_{{\rm G}}$, and ${\bf z}_{{\rm r},k}$  are uncorrelated with the estimated channel coefficients ${{\hat{\bf h}}_{{\rm d},k}}$, ${\hat{\bf G}}$ and ${{\hat{\bf h}}_{{\rm r},k}}$ \cite{Cheng2006TWCimperfectCSI,Taesang2006ITimperfectCSI,Dabbagh2008TcomimperfectCSI}.
Then, the true channel coefficients ${\bf h}_{{\rm d},k}$, $\bf G$ and ${\bf h}_{{\rm r},k}$ in the incoming data transmission frame can be modeled as a realization  from the sample space ${\mathcal F}\triangleq\{{\bf h}_{{\rm d},k}(\xi), {\bf G}(\xi),{{\bf h}_{{\rm r},k}}(\xi), \forall k, \forall \xi\}$ dominated by the knowledge of the imperfect CSI (${{\hat{\bf h}}_{{\rm d},k}}$, ${\hat{\bf G}}$ and ${{\hat{\bf h}}_{{\rm r},k}}$) and the distribution of the channel estimation error (${\bf z}_{{\rm d},k}$, ${\bf Z}_{{\rm G}}$, and ${\bf z}_{{\rm r},k}$), where $\xi$ denotes the index of the random realizations  drawn from ${\mathcal F}$.}\footnote{If the CSI is perfectly estimated, ${\bf h}_{{\rm d},k}(\xi)$, ${\bf G}(\xi)$, and ${\bf h}_{{\rm r},k}(\xi)$ keep constants for different $\xi$.}

%The instantaneous channel coefficients ${\bf h}_{{\rm d},k}$, $\bf G$ and ${\bf h}_{{\rm r},k}$ can be modeled as a realization  from the sample space ${\mathcal F}\triangleq\{{\bf h}_{{\rm d},k}(\xi), {\bf G}(\xi),{{\bf h}_{{\rm r},k}}(\xi), \forall k, \forall \xi\}$ dominated by the channel uncertainty,\footnote{If ${\bf h}_{{\rm d},k}(\xi)$, ${\bf G}(\xi)$, and ${\bf h}_{{\rm r},k}(\xi)$ keep constants for different $\xi$, there is no uncertainty, which means that the CSI is perfectly estimated.} where $\xi$ denotes the index of the random realizations  drawn from ${\mathcal F}$.

In this setup, the optimization problem is formulated as maximizing the expectation of the achievable WSR:
\begin{subequations}
\begin{align}
{\mathcal{P}}{\text{(B)}} \quad \max_{ {\bm \theta}} \quad & f_{\rm B}({\bm \theta} )= {\mathbb E}_{\xi} \left[\max_{ {\bm W}(\xi)} f_{\rm A}({\bf W}(\xi), {\bm \theta} ; \xi) \right]\notag \\
{\bf s.t.} \quad
& |\theta_n| =1, \quad \forall n=1,\cdots,N, \label{equ:P2c1}\\
& \sum_{k=1}^K \|{\bf w}_k (\xi) \|^2 \leq P_{\rm T}, \quad \forall \xi. \label{equ:P2c2}
%& \sum_k^K p_k \leq P_{\rm T}. \label{equ:P1c3}
\end{align}
\end{subequations}

Compared with existing works, the advantage of this setup is that, the RIS configuration only requires very small change from tradition wireless system.
However, ${\mathcal{P}}{\text{(B)}}$ is the stochastic optimization problem with inner-layer variable ${\bm W}(\xi)$ and outer-layer  variable $\bm \theta$, and both the inner-layer and outer-layer subproblems are non-convex with no closed-form solutions.
In addition, the objective function contains expectation operator, and  the probability density function of the sample space ${\mathcal F}$ usually very complicated with no closed-form expression as well.
Therefore, designing algorithm to solve ${\mathcal{P}}{\text{(B)}}$ is a really challenging task.

\section{Alternating Optimization for the Perfect CSI Setup}\label{Sec:AO_ideal}
The alternating optimization approach is the two-block version of the BCD method,
the basic idea of which is to decompose the optimization variables into several blocks, and then each block is updated following some specific rules while fixing the remaining blocks at their last updated values \cite{BCD2013updatingrule}.
In existing work \cite{zhangruiIRS} and \cite{YuenChauIRS}, the alternating optimization approach has been commonly used to address the joint optimization problem in the RIS-aided system.
In particular, the joint optimization problem is decomposed into two subproblems: one is the conventional beamforming design problem at the AP, and the other is the phase optimization problem given optimized beamforming vectors.

\subsection{Algorithm Description}\label{Sec:AO_ideal_algorithm}
When $\bm \theta$ is fixed, the subproblem to optimize $\bf W$ is reduced to the WSR maximization problem for the conventional multiuser MISO system.
This problem has been studied extensively in the literature, and one famous method to obtain the stationary solution is the WMMSE algorithm  with the following iterative updating rule \cite{WMMSE}:
\begin{subequations}
\begin{align}
\chi_k&=\left({\sum_{i=1}^K \left|{\bf h}_k^{\rm H} {\bf w}_i  \right|^2+\sigma_0^2}\right)^{-1} {\bf h}_k^{\rm H} {\bf w}_k, \label{equ:wmmse1}\\
\kappa_k&=\left(1-\chi_k^\ast {\bf h}_k^{\rm H} {\bf w}_k\right)^{-1}, \label{equ:wmmse2}\\
{\bf w}_k&=\omega_k \chi_k \kappa_k
\left({\lambda {\bf I}_M+\sum_{i=1}^K \omega_i |\chi_i|^2 \kappa_i {\bf h}_i {\bf h}_i^{\rm H} }\right)^{-1}
{   {\bf h}_k }, \label{equ:wmmse3}
\end{align}
\end{subequations}
where $\lambda \geq 0$ is the optimal dual variable for the transmit power constraint.

Then one can focus on the phase optimization subproblem.
For ease of representation, define the effective channels for the direct link and the RIS link as follows
\begin{subequations}
\begin{align}
{\bf a}_{i,k}&={\bf H}_{{\rm r},k} {\bf w}_i,\\
b_{i,k}&= {\bf h}_{{\rm d},k}^{\rm H} {\bf w}_i
,
\end{align}
\end{subequations}
respectively.
Then the phase optimization subproblem is represented as
\begin{align*}
{\mathcal{P}}{\text{(C)}} \; \max_{ {\bm \theta}} \; &
f_{\rm C}({\bm \theta})=\sum_{k=1}^K \omega_k \log (1+\frac{\left|{\bm \theta}^{\rm H} {\bf a}_{k,k}+b_{k,k}  \right|^2 }
{\sum_{i \neq k} \left|{\bm \theta}^{\rm H} {\bf a}_{i,k}+b_{i,k}  \right|^2+\sigma_0^2}) \notag \\
{\bf s.t.} \quad
& |\theta_n| =1, \quad \forall n=1,\cdots,N.
\end{align*}

One can see that, $f_{\rm C}({\bm \theta})$ is continuous and differentiable, and the constraint sets of $\bm \theta$ forms a complex circle manifold.
Thus, the stationary solution of ${\mathcal{P}}{\text{(C)}}$ can be obtained via the \emph{Riemannian conjugate gradient} (RCG) algorithm \cite{Manopt}, which has been widely applied for the analog precoder design in hybrid precoding problem \cite{Letaief2016manifoldHybrid} and has also shown good performance in the single-user RIS-aided MISO system \cite{Schober2019manifoldIRS}.
Conceptually, the RCG algorithm has three key steps in each iteration:
\subsubsection{Compute Riemannian Gradient}
The Riemannian gradient is the orthogonal projection of the Euclidean gradient $\nabla f_{\rm C}$ onto the complex circle:
\begin{equation*}
{\rm grad} f_{\rm C}= \nabla f_{\rm C}-{\rm Re} \left\{\nabla f_{\rm C} \circ {\bm \theta}^\ast  \right\}\circ {\bm \theta}
,
\end{equation*}
where the Euclidean gradient is
\begin{equation*}
\nabla f_{\rm C}=\sum_{k=1}^K 2\omega_k {\bf A}_k ,
\end{equation*}
with parameters
\begin{align*}
{\bf A}_k&=\frac{\sum_i {\bf a}_{i,k} {\bf a}_{i,k}^{\rm H}{\bm \theta}+ \sum_i {\bf a}_{i,k} b_{i,k}^\ast }
{\sum_{i} \left|{\bm \theta}^{\rm H} {\bf a}_{i,k}+b_{i,k}  \right|^2+\sigma_0^2}\\
  &\quad  -\frac{\sum_{i \neq k} {\bf a}_{i,k} {\bf a}_{i,k}^{\rm H}{\bm \theta}+ \sum_{i \neq k}  {\bf a}_{i,k} b_{i,k}^\ast }
  {\sum_{i \neq k} \left|{\bm \theta}^{\rm H} {\bf a}_{i,k}+b_{i,k}  \right|^2+\sigma_0^2}.
\end{align*}
\subsubsection{Search Direction}
One may find the tangent vector conjugate to ${\rm grad} f_{\rm C}$ as the search direction:
\begin{equation*}
{\bf d}=-{\rm grad} f_{\rm C}+\tau_1 {\mathcal T}({\bar{\bf d}})
,
\end{equation*}
where ${\mathcal T}(\cdot)$ is the vector transport function defined as
\begin{equation*}
{\mathcal T}({\bf d})={\bar{\bf d}}- {\rm Re} \left\{{\bf d}\circ {\bm \theta}^\ast \right\}\circ {\bm \theta}
,
\end{equation*}
$\tau_1$ is the conjugate gradient update parameter, and ${\bar{\bf d}}$ is the previous search direction.
\subsubsection{Retraction}
Project the tangent vector back to  the  complex circle manifold
\begin{equation*}
{\bm \theta}_n \leftarrow \frac{({\bm \theta}+\tau_2 {\bf d})_n} {|({\bm \theta}+\tau_2 {\bf d})_n|}
,
\end{equation*}
where $\tau_2$ is the Armijo step size.

%In this paper, we utilize the  ``Conjugate-gradient'' solver in the Manopt toolbox \cite{Manopt} to realize the RCG algorithm.

\subsection{Discussion}
The alternating optimization approach in Section \ref{Sec:AO_ideal_algorithm} is actually a multi-stage iterative optimization algorithm.
The outer loop involves two subproblems for optimizing $\bf W$ and $\bm \theta$, respectively, and each subproblem still requires iterative updating method to solve.

\textcolor{black}{
Specifically, the WMMSE algorithm requires the matrix inverse operations in all the three  updating steps with complexity ${\mathcal O}(K M^3)$.
In addition, in \eqref{equ:wmmse3}, one dimensional search (usually bi-search) for $\lambda$ is needed.
Hence, the complexity of the WMMSE algorithm is ${\mathcal O}(I_{\lambda} I_{\rm w} K M^3)$, where $I_{\lambda}$ and $I_{\rm w}$ are the iteration numbers of searching $\lambda$ and the three-step updating loop, respectively.
Besides, the complexity of the RCG algorithm is dominated by computing the Euclidean gradient, which is ${\mathcal O}(K^2 N^2)$.
The retraction step also requires iteratively searching $\tau_2$, but fortunately the complexity is only ${\mathcal O}(K^2 N)$ and can be ignored when $N$ is large.
Therefore, the total complexity of the alternating optimization approach is ${\mathcal O}\left(I_{\rm O}\left(I_{\lambda} I_{\rm w} K M^3+ I_{\rm R} K^2 N^2\right)\right)$, where $I_{\rm O}$ and $I_{\rm R}$ denote the iteration times of the outer loop, and the iteration times of the inner RCG algorithm, respectively.
}

The idea of alternatingly updating the variables is quite straightforward, and this method generally has good performance as verified by the simulations in \cite{zhangruiIRS} and \cite{YuenChauIRS}.
However,  %a new algorithm designing  is still necessary because:
this approach has two main drawbacks for the WSR problem in this paper:
%\begin{itemize}
%\item Firstly, it is known that the RIS has no capability to suppress inter-user interference. Hence, in contrast to the analog precoder design subproblem in hybrid precoding, $f_{\rm C}({\bm \theta})$ usually changes only a little after updating a new ${\bm \theta}$, and performance gain is mainly obtained by updating $\bf W$ given a new ${\bm \theta}$. Due to this deeply coupled relationship between ${\bm \theta}$ and $\bf W$, the converge speed of the outer-loop iteration is very slow, and usually we need large $I_{\rm R}$ to avoid the outer-loop iteration stopping in  uninteresting point, resulting in high  computational complexity. %Thus, it is unacceptable to process complex iterative algorithms to update ${\bm \theta}$ and $\bf W$ in each subproblem.
%\item Secondly, as the optimization algorithms for the two subproblems are designed independently, it is difficult to extend the  alternating optimization approach to the imperfect CSI setup, where the optimizing of $\bf W$ and $\bm \theta$ requires coordinately design.
%\end{itemize}
{\textcolor{black}{
\begin{itemize}
\item For the WSR maximization problem in the RIS-aided system,  the improvement of the objective function is mainly obtained by suppressing inter-user interference and power allocation among users, both of which are beyond the capabilities of the RIS.
    Thus, the beamforming design $\bf W$ and the phase vector $\bm \theta$ are deeply coupled. As a result, the  alternating optimization converges slowly, and its complexity becomes unacceptable when both the two subproblems also require iterative method to solve.
    In addition, for the iterative method on the phase optimization subproblem, the precision of the output $\bm \theta$ should be high enough (requiring large $I_{\rm R}$) to prevent the alternating optimization approach stopping at an uninteresting point.
\item If the two subproblems are solved independently, it is difficult to extend the algorithm to the imperfect CSI setup, where the optimization of $\bf W$ and $\bm \theta$ requires a coordinated design.
\end{itemize}
}
Therefore, a new algorithm designing with lower complexity and better extendibility is still necessary for ${\mathcal{P}}{\text{(A)}}$.

\section{Low-complexity BCD for the Perfect CSI Setup}\label{Sec:BCD_ideal}
In this section, we design new algorithm for ${\mathcal{P}}{\text{(A)}}$.
{\textcolor{black}{
To be specific, we first apply the closed-form FP approach in \cite{YuWei2018FP2} to  equivalently translate  the sum-of-logarithms-of-ratio problem ${\mathcal{P}}{\text{(A)}}$ into a more tractable form ${\mathcal{P}}{\text{(A1)}}$.
Then, we decompose ${\mathcal{P}}{\text{(A1)}}$ into four disjoint blocks.
Non-convex  BCD method \cite{nonconvexBCD2013LuoZQ} is exploited  to carry out the stationary solution for ${\mathcal{P}}{\text{(A1)}}$.
Specially, low-complexity updating rules for block ${\bf  W}$ and block ${\bm \theta}$ based on the  the prox-linear
BCD update rule \cite{BCD2013updatingrule} and the SCA method \cite{nonconvexBCD2013LuoZQ}, respectively.
}}

%the optimization variables are firstly decomposed into four disjoint blocks based on the FP technique \cite{YuWei2018FP2}.
%Compared with the two-block decomposition in Section \ref{Sec:AO_ideal}, the subproblem in each block in this section is more tractable.
%Then, new algorithm is designed by exploiting the recently proposed SCA technique based non-convex  BCD method \cite{nonconvexBCD2013LuoZQ}  to carry out the stationary solution for ${\mathcal{P}}{\text{(A)}}$.

%, in which the optimization variables can be well decomposed into multiple disjoint blocks, and then these blocks may be optimized coordinately based on the recently proposed non-convex  BCD method \cite{nonconvexBCD2013LuoZQ,nonconvexBCD2017YinWotao,BSUM2016LuoZQ}.

%Specifically, we first reformulate ${\mathcal{P}}{\text{(A)}}$ as a tractable form based on the fractional programming technique \cite{YuWei2018FP1,YuWei2018FP2}.
%Then,   surrogate functions are designed for subproblems to satisfy the convergence requirement of the non-convex BCD method \cite{nonconvexBCD2013LuoZQ}.
%Finally, a stationary solution for ${\mathcal{P}}{\text{(A)}}$ can be carried out by choosing one updating rule of BCD method.

\subsection{Closed-Form FP Approach}\label{Sec:Lagrangiandual}
The closed-form FP approach was proposed in \cite{YuWei2018FP2} to deal with the sum-of-logarithms-of-ratio problem as follows:
\begin{align*}
\max_{\bf x} \quad    \sum_{k=1}^K  \log(1+ \frac{|A_k({\bf x})|^2}{{ B_k}({\bf x})-|A_k({\bf x})|^2}),
\end{align*}
where ${{ B_k}({\bf x})}>|A_k({\bf x})|^2$ for all $k$.
Conceptually, the closed-form FP approach has two key steps:
\subsubsection{Lagrangian Dual Transform}
By introducing an auxiliary variable $\alpha_k$, the logarithm function can be tackled based on the following equation:
\begin{equation}\label{equ:add_alpha}
\log(1+\gamma_k)= \max_{\alpha_k \geq 0} \; \log(1+\alpha_k)-\alpha_k+\frac{ (1+\alpha_k) {\gamma}_k}{1+{\gamma}_k}.
\end{equation}
Then, the original problem is equivalently  transformed to
\begin{align}
\max_{{\bf x}, {\bm \alpha}} \quad & \sum_{k=1}^K \left(  \log(1+\alpha_k)- \alpha_k  +  (1+\alpha_k)\frac{|A_k({\bf x})|^2}{{ B_k}({\bf x})}\right) \notag \\
{\bf s.t.} \quad
& \alpha_k \geq 0, \quad \forall k=1,\cdots,K, \label{equ:Pc_alpha}
\end{align}
where ${\bm \alpha}=[\alpha_1, \cdots, \alpha_K]^{\rm T}$.
\subsubsection{Quadratic Transform}
Given ${\bm \alpha}$, one may focus on the following sum-of-ratios problem
\begin{align*}
\max_{\bf x} \quad    \sum_{k=1}^K \frac{|A_k({\bf x})|^2}{{ B_k}({\bf x})}.
\end{align*}
The key idea is introducing auxiliary variables ${\bm \beta}=[\beta_1, \cdots, \beta_K]^{\rm T}$, and then the above problem is  equivalently translated to
\begin{align*}
\max_{{\bf x}, {\bm \beta}} \quad    \sum_{k=1}^K
\left( 2 {\rm Re} \left\{ \beta_k^{\ast} A_k({\bf x})\right\}- \left|\beta_k \right|^2 { B_k}({\bf x})
\right).
\end{align*}
The equivalence can be verified by substituting $\beta_k = \frac{ A_k({\bf x}) } {{ B_k}({\bf x})}$ into above problem.

\subsection{Non-Convex BCD}
Applying the closed-form FP approach introduced  above, problem ${\mathcal{P}}{\text{(A)}}$  is equivalent to following problem
\begin{align*}
{\mathcal{P}}{\text{(A1)}} \; \max_{ {\bm \alpha},{\bm \beta}, {\bf W}, {\bm \theta}} \quad &   f_{\text A1}({\bm \alpha},{\bm \beta},{\bf  W}, {\bm \theta})  \\
{\bf s.t.} \quad
& \eqref{equ:P1c1}, \eqref{equ:P1c2}, \eqref{equ:Pc_alpha},
\end{align*}
where the new objective function is
\begin{equation*}
\begin{aligned}[b]
&f_{\text A1}({\bm \alpha},{\bm \beta},{\bf  W}, {\bm \theta})
%&=\sum_{k=1}^K \omega_k \log(1+\alpha_k)-\sum_{k=1}^K \omega_k \alpha_k \\
%&\qquad + f_{\text{A3}}({\bf  W}, {\bm \theta},{\bm \beta})\\
=\sum_{k=1}^K \omega_k \left(\log\left(1+\alpha_k\right)- \alpha_k \right)\\
&\qquad+\sum_{k=1}^K 2 \sqrt{\omega_k (1+\alpha_k)}  {\rm Re} \left\{ \beta_k^{\ast} ({\bf h}_{{\rm d},k}^{\rm H}+{\bm \theta}^{\rm H} {\bf H}_{{\rm r},k}) {\bf w}_k \right\}\\
&\qquad-\sum_{k=1}^K \left|\beta_k \right|^2 \left({\sum_{i=1}^K \left|({\bf h}_{{\rm d},k}^{\rm H}+{\bm \theta}^{\rm H} {\bf H}_{{\rm r},k}){\bf w}_i  \right|^2+\sigma_0^2}\right).
\end{aligned}
\end{equation*}
In this paper, we adopt the BCD method \cite{BCD2013updatingrule} to decompose the optimization variables ${\bm \alpha}$, $ {\bm \beta}$, ${\bf  W}$, and ${\bm \theta}$, and aim to get a stationary solution for ${\mathcal{P}}{\text{(A1)}}$.

The BCD is an iterative method, where ${\bm \alpha}$, ${\bm \beta}$, ${\bf  W}$, and ${\bm \theta}$ are cyclically updated.
To be specific, denote by $\bar{\bm \alpha}$, $\bar{\bm \beta}$, $\bar{\bf  W}$, and $\bar{\bm \theta}$ the temporal optimization results in last iteration.
Then, it is easy to carry out
%\begin{subequations}
\begin{align}
{\alpha}_k&=\frac{{\bar{\zeta}}_k^2+{\bar{\zeta}}_k\sqrt{{\bar{\zeta}}_k^2+4}}{2}, \label{equ:A_update_alpha}\\
{\beta}_k&=\frac{ \sqrt{\omega_k (1+{\bar{\alpha}}_k) } ({\bf h}_{{\rm d},k}^{\rm H}+{\bar{\bm \theta}}^{\rm H} {\bf H}_{{\rm r},k}) {\bar{\bf w}}_k }
{\sum_{i=1}^K \left|({\bf h}_{{\rm d},k}^{\rm H}+{\bar{\bm \theta}}^{\rm H} {\bf H}_{{\rm r},k}){\bar{\bf w}}_i  \right|^2+\sigma_0^2}, \label{equ:A_update_beta}
\end{align}
%\end{subequations}
where ${\bar{\zeta}}_k=\frac{1}{\sqrt{\omega_k}}{\rm Re} \left\{ {\bar{\beta}}_k^{\ast} {\bar{\bf h}}_k^{\rm H} {\bar{\bf w}}_k \right\}$ and ${\bar{\bf h}}_k$ the combined channel:
\begin{equation*}
{\bar{\bf h}}_k={\bf h}_{{\rm d},k}+{\bf H}_{{\rm r},k}^{\rm H} {\bar{\bm \theta}}.
\end{equation*}
Remark that, in \cite{YuWei2018FP2}, it is suggested updating ${\alpha}_k$ by the  temporal SINR which is not BCD. Nevertheless, its convergence is established as well \cite[Appendix A]{YuWei2018FP2}.

\subsection{Prox-linear Update for ${\bf  W}$}
One may update ${\bf  W}$ by solving following problem:
\begin{align*}
{{\bm W}}=\arg & \max_{{\bm W}}  \;  f_{\text A2}({{\bf  W}}),
 \\
{\bf s.t.} \quad
& \sum \|{\bf w}_k\|^2 \leq P_{\rm T},
\end{align*}
where $f_{\text A2}({{\bf  W}})=f_{\text A1}({\bar{\bm \alpha}},{\bar{\bm \beta}},{{\bf  W}}, {\bar{\bm \theta}})$,
and thus have:
\begin{equation*}
{\bm w}_k=\sqrt{\omega_k (1+{\bar{\alpha}}_k) } {\bar{\beta}}_k
\bigg(\lambda {\bf I}_M   +\sum_{i=1}^K |{\bar{\beta}}_i|^2 {\bar{\bf h}}_i {\bar{\bf h}}_i^{\rm H} \bigg)^{-1}
{   {\bar{\bf h}}_k }, \label{equ:A_update_W}
\end{equation*}
where $\lambda \geq 0$ is the optimal dual variable for the transmit power constraint.
However, the matrix inverse operation is expensive, and to obtain a high-accurate ${\bf  W}$, the iteration numbers for searching $\lambda$ is usually high.

To eliminate the one dimensional search of $\lambda$ as well as the expensive matrix inverse operation, we apply the prox-linear BCD update rule as follows \cite{BCD2013updatingrule}:
\begin{align*}
{{\bm W}}=\arg  & \min_{{\bm W}} \sum_{k=1}^K \left({\rm Re} \left\{ {\bf g}_k^{\rm H}({\bf w}_k-{\hat{\bf w}}_k)\right\}+\frac{L}{2} \|{\bf w}_k-{\hat{\bf w}}_k \|^2 \right)
\\
{\bf s.t.} \quad
& \sum \|{\bf w}_k\|^2 \leq P_{\rm T},
\end{align*}
where $L>0$,
the gradient is denoted by
\begin{align*}
{\bf g}_k&=-{\left.\frac{\partial f_{\text A2}}{\partial {\bf w}_k}\right|}_{{\bf w}_k={\hat{\bf w}}_k}  \\
&=-2 \sqrt{\omega_k (1+{\bar{\alpha}}_k)}   {\bar \beta}_k {\bar{\bf h}}_k
+2 \sum_{i=1}^K |{\bar{\beta}}_i|^2 {\bar{\bf h}}_i {\bar{\bf h}}_i^{\rm H} {\hat{\bf w}}_k,
\end{align*}
${\hat{\bf w}}_k={\bar{\bf w}}_k+\epsilon \left({\bar{\bf w}}_k-{\ddot{\bf w}}_k\right)$ is the extrapolated point, ${\ddot{\bf w}}_k$ is the value of ${\bf w}_k$ before it was updated to ${\bar{\bf w}}_k$,
and $\epsilon \geq 0$ is the extrapolation weight.
Then we have a simple update rule:
\begin{align}
{\bm w}_k&= \frac{1}{L-2 \lambda} \left(L {\hat{\bf w}}_k- {\bf g}_k\right), \label{equ:A_update_w}\\
\lambda &= \frac{L}{2}- \frac{1}{2 P_{\rm T}} \sum_{k=1}^k \left\|L {\hat{\bf w}}_k- {\bf g}_k\right\|^2.
\end{align}
One can see that, the complexity to update ${{\bm W}}$ is reduced to ${\mathcal O}(K M^2)$, and no iteration is required.

We set $L=2 \left\| \sum_{i=1}^K |{\bar{\beta}}_i|^2 {\bar{\bf h}}_i {\bar{\bf h}}_i^{\rm H}\right\|_{\rm F}$, which is the Lipschitz constant of the gradient ${\bf g}_k$.
Then, the  extrapolation weight is taken by
\begin{equation*}
\epsilon= \min \left(\frac{d-1}{\bar d}, 0.9999 \sqrt{\frac{\bar L}{L}} \right),
\end{equation*}
where $\bar d$ and $\bar L$ are the values adopted in previous iteration, and $d$ is  recursively defined by
$d=\frac{1}{2}\left(1+\sqrt{1+4 {\bar d}^2}\right)$ with initial value $1$.
%$d_0=1$ and $d_t=\frac{1}{2}\left(1+\sqrt{1+4 d_{t-1}^2}\right)$, where $t$ indicates the index of iteration step.
Since $f_{\text A2}({{\bf  W}})$ is strongly convex which satisfies the KL property \cite{BCD2013updatingrule},  the convergence of the prox-linear update rule is established (see \cite[Lemma 2.2]{BCD2013updatingrule}).

\subsection{Successive Convex Approximation for Updating $\bm \theta$}\label{Sec:sca_perfect}
In conventional BCD method, ${\bm \theta}$ is updated according to ${\bm \theta}=\arg \max_{{\bm \theta}} f_{\text A1}({\bar{\bm \alpha}},{\bar{\bm \beta}},{\bar{\bf  W}}, {{\bm \theta}})$.
After dropping irrelevant constant terms, this updating rule is represented as
\begin{align*}
{\bm \theta}= \arg \min_{{ {\bm \theta}}} &
\; f_{\text A3}({ {\bm \theta}})\triangleq{\bm \theta}^{\rm H}  {\bm U} {\bm \theta}
- 2 {\rm Re} \left\{ {\bm \theta}^{\rm H} {\bm \nu}
\right\} \\
{\bf s.t.} \quad
& |\theta_n| =1, \quad \forall n=1,\cdots,N.
\end{align*}
where ${\bm U}$ and $\bm \nu$ are
\begin{subequations}
\begin{align}
{\bm U}&=
\sum_{k=1}^K \left|{\bar{\beta}}_k \right|^2
\sum_{i=1}^K  {\bar {\bf a}}_{i,k} {\bar {\bf a}}_{i,k}^{\rm H}, \label{equ:A_U}
 \\
%{\bm \nu}&= \sum_{k=1}^K    \sqrt{\omega_k (1+{\bar \alpha}_k) } {\bar \beta}_k^{\ast} {\bf H}_{{\rm r},k} {\bar{\bf w}}_k. \label{equ:A_v}\\
{\bm \nu}&= \sum_{k=1}^K  \left(  \sqrt{\omega_k (1+{\bar \alpha}_k)} {\bar \beta}_k^{\ast} {\bar {\bf a}}_{k,k}
    -  \left|\beta_k\right|^2 \sum_{i=1}^K  {\bar b}_{i,k}^{\ast}  {\bar {\bf a}}_{i,k}
\right), \label{equ:A_v}
\end{align}
\end{subequations}
with ${\bar {\bf a}}_{i,k}={\bf H}_{{\rm r},k} {\bar{\bf w}}_i$, and ${\bar b}_{i,k}= {\bf h}_{{\rm d},k}^{\rm H} {\bar{\bf w}}_i$.
We further replace $\theta_n$ by $\varphi_n$, where $\theta_n= e^{\jmath \varphi_n}$ and $\varphi_n \in {\mathbb R}$, and then the update rule is recast to
\begin{align*}
 {\bm \varphi}= \arg \min_{{ {\bm \varphi}\in {\mathbb R}^N}} &
\; f_{\text A4}({ {\bm \varphi}}) \triangleq ({e^{\jmath {\bm \varphi} }})^{\rm H}  {\bm U} {e^{\jmath {\bm \varphi} }}
- 2 {\rm Re} \left\{ {\bm \nu}^{\rm H} {e^{\jmath {\bm \varphi} }}
\right\}
\end{align*}
where ${\bm \varphi}=[\varphi_1, \cdots, \varphi_N]^{\rm T}$.

{\color{black}{
However, $f_{\text A4}({ {\bm \varphi}})$ is non-convex, and its really hard to solve the optimal solution.
Fortunately, it is pointed out in \cite{nonconvexBCD2013LuoZQ} that, we only need to solve the following surrogate problem by exploiting the SCA technique, and the BCD method will still converge to a stationary solution.
In particular, denote the  surrogate function for $f_{\text A4}({ {\bm \varphi}})$ by $f_{\text A5}({ {\bm \varphi}}, { \bar{\bm \varphi}})$, and the output ${\bm \varphi}$ is obtained from
\begin{equation*}
\begin{aligned}[b]
 {\bm \varphi}= \arg \min_{{ {\bm \varphi}\in {\mathbb R}^N}} &
\; f_{\text A5}({ {\bm \varphi}}, { \bar{\bm \varphi}}).
\end{aligned}
\end{equation*}
It is known that $f_{\text A4}({ {\bm \varphi}})$ continuously differentiable, and thus we need $f_{\text A5}({ {\bm \varphi}}, { \bar{\bm \varphi}})$ satisfies following two constraint \cite[Proposition 1]{nonconvexBCD2013LuoZQ}:
\begin{subequations}
\begin{align}
f_{\text A5}({ \bar{\bm \varphi}}, { \bar{\bm \varphi}}) &= f_{\text A4}({ \bar{\bm \varphi}}),\label{equ:sca_1}\\
f_{\text A5}({ {\bm \varphi}}, { \bar{\bm \varphi}}) &\geq f_{\text A4}({ {\bm \varphi}}).\label{equ:sca_2}
\end{align}
\end{subequations}
}}

{\color{black}{
In this paper, we adopt the surrogate function constructed by the second order Taylor expansion \cite{MM2017Palomar}:
\begin{equation*}
\begin{aligned}[b]
f_{\text A5}({ {\bm \varphi}}, { \bar{\bm \varphi}}) = f_{\text A4}({ \bar{\bm \varphi}})+\nabla f_{\text A4}({ \bar{\bm \varphi}})^{\rm T}({\bm \varphi}-{{ \bar{\bm \varphi}}})+\frac{\kappa}{2}\|{\bm \varphi}-{{ \bar{\bm \varphi}}}\|^2,
\end{aligned}
\end{equation*}
where $\nabla f_{\text A4}({ \bar{\bm \varphi}})=2 {\rm Re} \left\{ -\jmath {\bar{\bm \theta}}^\ast \circ  \left({\bm U} {\bar{\bm \theta}}- {\bm \nu}\right) \right\}$ is the gradient, and $\kappa$ is chosen to satisfy \eqref{equ:sca_2}.
%make sure that the updated ${\bm \varphi}$ may improve the original objective function:
Finally, ${\bm \varphi}$ is updated by the minimum value of $f_{\text A5}({ {\bm \varphi}}, { \bar{\bm \varphi}})$:
\begin{equation}\label{equ:A_update_varphi}
{\bm \varphi}={\bar{\bm \varphi}}-\frac{\nabla f_{\text A4}({\bar{\bm \varphi}})}{\kappa}.
\end{equation}
}}

\subsection{Algorithm Development}\label{Sec:development_A}
%We have designed the update rules for all the four blocks with closed-form solution except the step size $\kappa$ in \eqref{equ:A_update_varphi}.
The block selection rule for BCD method is designed as follows
\begin{equation}\label{equ:BCD_select}
\cdots\cdots
 {\bm \alpha}\rightarrow {\bm \beta}\rightarrow{\bm \varphi} \rightarrow {\bm \beta} \rightarrow{\bf W} \rightarrow {\bm \alpha} \cdots \cdots
\end{equation}
The stationary solution for ${\mathcal{P}}{\text{(A)}}$ can be carried out by simply setting $\kappa$ as the Lipschitz constant of $\nabla f_{\text A4}$ \cite{nonconvexBCD2013LuoZQ}. It is known that, the complexity to update ${\bm \alpha}$, ${\bm \beta}$, and ${\bf W}$ are ${\mathcal O}(K N M)$, ${\mathcal O}(K N M)$, and ${\mathcal O}(K M^2)$, respectively. Besides,  the complexity of update ${\bm \varphi}$ is dominated by the parameter $\bf U$ in \eqref{equ:A_U}, which is ${\mathcal O}(K^2 N^2)$.
Therefore, the total complexity of the proposed BCD method is ${\mathcal O}\left(I_{\rm O}\left(2 K N M+K M^2+  K^2 N^2\right)\right)$, where $I_{\rm O}$ is the number of iterations.

However, the improvement of $f_{\text A3}$ by updating ${\bm \varphi}$ is even much smaller than that of $f_{\rm C}$ in the alternating optimization approach in Section \ref{Sec:AO_ideal}. So the convergence speed is much slower, and the algorithm complexity may not decrease.
In the next, we show that the convergence of the proposed BCD algorithm can be accelerated by chosen a proper search step size in \eqref{equ:A_update_varphi}.
%To accelerate the convergence of BCD, one straightforward method is  iteratively taking \eqref{equ:A_update_varphi} to update ${\bm \varphi}$ until $f_{\text A3}$ converges. Since the complexity of $\nabla f_{\text A4}({\bar{\bm \varphi}})$ is ${\mathcal O}(N^2)$, the total complexity of  BCD becomes ${\mathcal O}\left(I_{\rm O}\left(2 K N M+K M^2+  K^2 N^2+I_{\bm \varphi}N^2\right)\right)$, where $I_{\bm \varphi}$ is the iteration number of the inner loop for updating ${\bm \varphi}$. Hence, compared with the alternating optimization approach, the complexity about $N$ is reduced by $K^2$ times.
%In the next, we will show that the complexity of the BCD algorithm in this section can be further reduced by designing a proper $\kappa$, meanwhile the convergence speed is still guaranteed.

\begin{algorithm}[!t]
\caption{$[{\bm U},{\bm \nu}]={\text{ParfunA}({\bm \varphi},\bar{\bm \alpha}, \bar{\bm \beta}, \bar{\bf  W})} $.}
\label{alg:ParfumA1}
\begin{algorithmic}[1]
%\REQUIRE
% channel coefficients ${\bf G}$, ${\bf h}_{{\rm d},k}$ and ${\bf h}_{{\rm r},k}$ for all $k$,  and the transmit power constraint $P_{\rm T}$.
\STATE Update ${\bm \beta}$ by \eqref{equ:A_update_beta};\\
\STATE Update ${\bm W}$ by \eqref{equ:A_update_w};\\
\STATE Update ${\bm \alpha}$ by \eqref{equ:A_update_alpha};\\
\STATE Update ${\bm \beta}$ based on \eqref{equ:A_update_beta};\\
\STATE Update ${\bm U}$ and ${\bm \nu}$ by \eqref{equ:A_U} and \eqref{equ:A_v}, respectively.
\end{algorithmic}
\end{algorithm}

Consider following problem for ${\bm \varphi}$
\begin{align*}
{\mathcal{P}}{\text{(A6)}} \; \min_{{ {\bm \varphi}}} &
\; f_{\text A6}({ {\bm \varphi}})\triangleq ({e^{\jmath {\bm \varphi} }})^{\rm H}  {\bm U} {e^{\jmath {\bm \varphi} }}
- 2 {\rm Re} \left\{ {\bm \nu}^{\rm H} {e^{\jmath {\bm \varphi} }}
\right\}
\end{align*}
where $[{\bm U},{\bm \nu}]={\text{ParfunA}({\bm \varphi},\bar{\bm \alpha}, \bar{\bm \beta}, \bar{\bf  W})} $ which is summarized in Algorithm \ref{alg:ParfumA1}.
According to the block selection rule in \eqref{equ:BCD_select}, every stationary solution of ${\mathcal{P}}{\text{(A6)}}$ is the critical point of the BCD method, which is as well the  stationary solution of ${\mathcal{P}}{\text{(A)}}$.

To solve ${\mathcal{P}}{\text{(A6)}}$, let's construct function as follows
\begin{equation}\label{equ:Surrogate_fun_A6}
\begin{aligned}[b]
h_{\text A6}({ {\bm \varphi}}, { \bar{\bm \varphi}}) = f_{\text A6}({ \bar{\bm \varphi}})+\nabla f_{\text A4}({ \bar{\bm \varphi}})^{\rm T}({\bm \varphi}-{{ \bar{\bm \varphi}}})+\frac{\kappa}{2}\|{\bm \varphi}-{{ \bar{\bm \varphi}}}\|^2,
\end{aligned}
\end{equation}
where the parameters ${\bm U}$ and ${\bm \nu}$ in $\nabla f_{\text A4}({ \bar{\bm \varphi}})$ are determined by $[{\bm U},{\bm \nu}]={\text{ParfunA}(\bar{\bm \varphi},\bar{\bm \alpha}, \bar{\bm \beta}, \bar{\bf  W})} $.
Then,  one can verify that, for sufficient large $\kappa$, we always have
%\begin{subequations}
\begin{align*}
h_{\text A6}({ \bar{\bm \varphi}}, { \bar{\bm \varphi}}) &= f_{\text A6}({ \bar{\bm \varphi}}),\\
h_{\text A6}({ {\bm \varphi}}, { \bar{\bm \varphi}}) &\geq f_{\text A6}({ {\bm \varphi}}).
\end{align*}
%\end{subequations}
It is known that, both $f_{\text A5}$ and $f_{\text A6}$ are continuously differentiable functions, and the constraints in \eqref{equ:sca_1} and \eqref{equ:sca_2} are satisfied. Therefore, $h_{\text A6}({ {\bm \varphi}}, { \bar{\bm \varphi}})$ is the SCA  surrogate function of $f_{\text A6}({ {\bm \varphi}})$, and we shall have the gradient of $f_{\text A6}({ {\bm \varphi}})$:
\begin{equation}\label{equ:equation_gradient}
\begin{aligned}[b]
\nabla f_{\text A6}({ {\bm \varphi}})&=\nabla f_{\text A4}({ {\bm \varphi}})\\
&=2 {\rm Re} \left\{ -\jmath {e^{-\jmath {\bm \varphi} }} \circ  \left({\bm U} {{e^{\jmath {\bm \varphi} }}}- {\bm \nu}\right) \right\},
\end{aligned}
\end{equation}
since both $f_{\text A5}$ and $f_{\text A6}$ are continuously differentiable functions (see \cite[Proposition 1]{nonconvexBCD2013LuoZQ}).

From \eqref{equ:equation_gradient}, the update rule in \eqref{equ:A_update_varphi} has the same formation as the gradient projection algorithm for ${\mathcal{P}}{\text{(A6)}}$.
We design the step size $\kappa$ can be determined by the Armijo rule \cite{Nonlinearprogram}:
\begin{equation}\label{equ:A_update_varphi_armijo}
f_{\text A6}({ \bar{\bm \varphi}})-f_{\text A6}({ {\bm \varphi}}) \geq \zeta \kappa \|\nabla f_{\text A4}({\bar{\bm \varphi}})\|_2^2,
\end{equation}
where $0<\zeta<0.5$, $\kappa$ is the largest element in $\{\kappa_0^{-j}\}_{j=0,1,\dots}$ and $\kappa_0>1$.
%
% which satisfies
%\begin{equation}\label{equ:A_update_varphi_con}
%f_{\text A1}({\bar{\bm \alpha}},\bar{\bm \beta},\bar{\bf  W}, {\bm \varphi})>f_{\text A1}({\bar{\bm \alpha}},\bar{\bm \beta},\bar{\bf  W}, \bar{\bm \varphi}),
%\end{equation}
%
%Since in every iteration, the value of
%Obviously, $\kappa$ designed by \eqref{equ:A_update_varphi_armijo} satisfies the updating constraint in \eqref{equ:A_update_varphi_con}, and thus the BCD method will converge.

\textcolor{black}{
The proposed BCD algorithm above is summarized in Algorithm \ref{alg:P1}.
Denote by $I_{\rm A}$ the iteration number of the Armijo search.
The complexity of  the proposed algorithm is ${\mathcal O}\left(I_{\rm O}\left(I_{\rm A}\left(2 K N M+K M^2\right)+  K^2 N^2\right)\right)$.
We will show in simulation that, the proposed algorithm has nearly the same $I_{\rm O}$ as that of the alternating optimization approach in Section \ref{Sec:AO_ideal}.
Hence, the complexity of Algorithm \ref{alg:P1} with respect to $N$ is reduced by $I_{\rm R}$ times, since no iterative updating is required in the block with respect to $\bm \theta$.
%One can see that, compared with the alternating optimization approach in Section \ref{Sec:AO_ideal}, the complexity about $N$ is reduced by $I_{\rm R}$ times (no iterative updating is needed for this subproblem).
}

\begin{algorithm}[!t]
\caption{Low-complexity BCD for the perfect CSI setup.}
\label{alg:P1}
\begin{algorithmic}[1]
%\REQUIRE
% channel coefficients ${\bf G}$, ${\bf h}_{{\rm d},k}$ and ${\bf h}_{{\rm r},k}$ for all $k$,  and the transmit power constraint $P_{\rm T}$.
\STATE {Initialize ${ {\bm W}}^{(0)}$ and ${\bm \theta}^{(0)}$ to feasible values.\\
\STATE Initialize ${\bm \alpha}^{(0)}$ and ${\bm \beta}^{(0)}$ by \eqref{equ:A_update_alpha} and \eqref{equ:A_update_beta}, and set $t=0$.
{\bf Repeat}}
\STATE  Design $\kappa$ using \eqref{equ:A_update_varphi_armijo} according to ${\bm \varphi}^{(t)}$ , ${\bm \alpha}^{(t)}$, ${\bm \beta}^{(t)}$, ${\bf  W}^{(t)}$;
\STATE Set $t=t+1$;
\STATE  Update ${\bm \varphi}^{(t)}$ , ${\bm \alpha}^{(t)}$, ${\bm \beta}^{(t)}$, ${\bf  W}^{(t)}$.
\\
{\bf Until} The value of the objective function $f_{\text A1}$ converges.
\end{algorithmic}
\end{algorithm}

\section{Extend the Non-convex BCD for the Imperfect CSI Setup}\label{Sec:Imperfect}
In this section, we extend the non-convex BCD method in Algorithm \ref{alg:P1} to solve ${\mathcal{P}}{\text{(B)}}$.
Specifically, after applying the closed-form FP approach in Section \ref{Sec:Lagrangiandual}, ${\mathcal{P}}{\text{(B)}}$  is equivalently transformed as follows:
%\begin{subequations}
\begin{align*}
{\mathcal{P}}{\text{(B1)}} \quad \min_{ {\bm \theta}} \quad & f_{\rm B1}({\bm \theta} )={\mathbb E}_{\xi}
\left[ g({\bm \theta}; \xi ) \right]\notag \\
{\bf s.t.} \quad
& |\theta_n| =1, \quad \forall n=1,\cdots,N, %\label{equ:Pbc1}
%& \sum_{k=1}^K \|{\bf w}_k (\xi)\|^2 \leq P_{\rm T}, \quad \forall \xi. \label{equ:Pbc2}
%& \sum_k^K p_k \leq P_{\rm T}. \label{equ:P1c3}
\end{align*}
%\end{subequations}
where
%\begin{subequations}
\begin{align*}
g({\bm \theta}; \xi ) =
  \min_{ {\bm \alpha},{\bm \beta}, {\bf W}} & \; -f_{\text A1}({\bm \alpha},{\bm \beta},{\bf  W}, {\bm \theta}; \xi) \notag
 \\
 {\bf s.t.} \quad
& \sum_{k=1}^K \|{\bf w}_k \|^2 \leq P_{\rm T}, \\%\label{equ:fbc1}\\
& \alpha_k \geq 0, \quad  \; \forall k=1,\cdots,K. %\label{equ:fbc2}
\end{align*}
%\end{subequations}

\subsection{Solve The Inner-Layer Subproblem}
Exploiting the BCD approach in Section \ref{Sec:BCD_ideal}, we may obtain a stationary solution for $g({\bm \theta}; \xi )$, and have following approximated function:
\begin{equation*}
{\hat g}({\bm \theta}; \xi ) =
{\bm \theta}^{\rm H}  {\bm U} {\bm \theta}
- 2 {\rm Re} \left\{ {\bm \nu}^{\rm H} {\bm \theta}
\right\},
\end{equation*}
where $[{\bm U},{\bm \nu}]={\text{ParfunB}({\bm \varphi}, {\bm \alpha}^{(0)}, {\bm \beta}^{(0)}, {\bf  W}^{(0)}; \xi)} $ which is summarized in Algorithm \ref{alg:ParfumB}, and  ${\bm \varphi}=[\varphi_1, \cdots, \varphi_N]^{\rm T}$ with $\theta_n= e^{\jmath \varphi_n}$.
One can also verify that, ${\hat g}({\bm \theta}; \xi )$ is  continuously differentiable, which has the unique and finite output given fixed ${\bm \alpha}^{(0)}$, ${\bm \beta}^{(0)}$, ${\bf  W}^{(0)}$, and stop criterion.
Then ${\mathcal{P}}{\text{(B1)}}$ can be approximately solved by
\begin{align*}
{\mathcal{P}}{\text{(B2)}} \quad \min_{ {\bm \varphi}} \; & f_{\rm B2}({\bm \varphi} )={\mathbb E}_{\xi}
\left[ {\hat g}({\bm \varphi}; \xi ) \right]
\end{align*}
where
\begin{equation*}
{\hat g}({\bm \varphi}; \xi ) =
({e^{\jmath {\bm \varphi} }})^{\rm H}  {\bm U} {e^{\jmath {\bm \varphi} }}
- 2 {\rm Re} \left\{ {\bm \nu}^{\rm H} {e^{\jmath {\bm \varphi} }}
\right\}.
\end{equation*}
Since we obtain the stationary solution for the inner-layer subproblem, the stationary solution of ${\mathcal{P}}{\text{(B2)}}$ is also the stationary solution of ${\mathcal{P}}{\text{(B1)}}$.

\begin{algorithm}[!t]
\caption{$[{\bm U},{\bm \nu}]={\text{ParfunB}({\bm \varphi}, {\bm \alpha}^{(0)}, {\bm \beta}^{(0)}, {\bf  W}^{(0)}; \xi)} $.}
\label{alg:ParfumB}
\begin{algorithmic}[1]
\REQUIRE
Channel realizations ${\bf h}_{{\rm d},k}(\xi)$, ${\bf G}(\xi)$, ${{\bf h}_{{\rm r},k}}(\xi)$ for all $k$;
\STATE {Set $t=0$.\\
{\bf Repeat}}
\STATE  Design $\kappa$ using \eqref{equ:A_update_varphi_armijo} according to ${\bm \varphi}$ , ${\bm \alpha}^{(t)}$, ${\bm \beta}^{(t)}$, ${\bf  W}^{(t)}$;
\STATE Set $t=t+1$;
\STATE  Update  ${\bm \alpha}^{(t)}$, ${\bm \beta}^{(t)}$, ${\bf  W}^{(t)}$.
\\
{\bf Until} The value of the objective function $f_{\text A1}$ converges.
\STATE Update ${\bm U}$ and ${\bm \nu}$ by \eqref{equ:A_U} and \eqref{equ:A_v}, respectively.
\end{algorithmic}
\end{algorithm}

\subsection{Stochastic Successive Convex Approximation for ${\mathcal{P}}{\text{(B2)}}$}
A classical approach to deal with the expectation operation in $f_{\rm B2}({\bm \varphi} )$ is the sample average approximation method.
To be specific, at the $r$-th iteration, a new realization $\xi_r$ is obtained and ${\bm \varphi}$ is updated by:
\begin{equation}\label{equ:saa}
{\bm \varphi}_r= \arg \; \min_{ {\bm \varphi}} \; f_{{\rm B},r}({\bm \varphi})\triangleq \frac{1}{r} \sum_{i=1}^r {\hat g}({\bm \varphi}; \xi_i ).
\end{equation}
\subsubsection{Stochastic SCA}
However, ${\hat g}({\bm \varphi}; \xi_i )$ is still a non-convex function of ${\bm \varphi}$, and the parameters ${\bm U}$ and ${\bm \nu}$ are obtained by iterative operations.
In \cite{ZQLuo2016SSUM}, a stochastic optimization version of SCA is proposed to deal with this kind of problem, in which  ${\bm \varphi}$ is updated by:
\begin{equation}\label{equ:ssca-p1}
{\bm \varphi}_r=\arg \; \min_{ {\bm \varphi}} \; h_r({\bm \varphi})\triangleq\frac{1}{r} \sum_{i=1}^r {\hat h}_i({\bm \varphi}, {\bm \varphi}_{i-1}; \xi_i ),
\end{equation}
where ${\hat h}_i({\bm \varphi}, {\bm \varphi}_{i-1}; \xi_i )$ is an approximation of ${\hat g}({\bm \varphi}; \xi_i )$ around the output of $\{i-1\}$-th iteration.

Similar to the conventional SCA constraints in \eqref{equ:sca_1} and \eqref{equ:sca_2}, if ${\hat h}_i({\bm \varphi}_1, {\bm \varphi}_2; \xi_i )$ is continuously differentiable strongly convex with uniformly bounded second order derivatives, we still require following two constraints to guarantee the convergence of the  stochastic SCA:
\begin{subequations}
\begin{align}
{\hat h}_i({\bm \varphi}, {\bm \varphi}; \xi_i ) &= {\hat g}({\bm \varphi}; \xi_i ), \label{equ:ssca_c1}\\
{\hat h}_i({\bm \varphi}_1, {\bm \varphi}_2; \xi_i ) &\geq {\hat g}({\bm \varphi}_1; \xi_i ). \label{equ:ssca_c2}
\end{align}
\end{subequations}
\subsubsection{Design Surrogate Function}
Fortunately, the  surrogate function designed for the perfect CSI setup in \eqref{equ:Surrogate_fun_A6} also satisfies above requirement. Thus we have
\begin{equation}\label{equ:Surrogate_fun}
\begin{aligned}[b]
{\hat h}_i({ {\bm \varphi}}, { {\bm \varphi}_{i-1}}; \xi_i) &= {\hat g}({\bm \varphi}_{i-1}; \xi_i )+\frac{\kappa_i}{2}\|{\bm \varphi}-{\bm \varphi}_{i-1}\|^2\\
&\quad +\nabla {\hat g}({\bm \varphi}_{i-1}; \xi_i )^{\rm T}({\bm \varphi}-{\bm \varphi}_{i-1}),
\end{aligned}
\end{equation}
where the gradient is
%\begin{equation*}
\begin{subequations}
\begin{align}
\nabla {\hat g}({\bm \varphi}_{i-1}; \xi_i )
&=2 {\rm Re} \left\{ -\jmath {e^{-\jmath {\bm \varphi}_{i-1} }} \circ  \left(\bar{\bm U} {{e^{\jmath {\bm \varphi}_{i-1} }}}- \bar{\bm \nu}\right) \right\}, \label{equ:ssca_grad1}\\
[\bar{\bm U},\bar{\bm \nu}]&={\text{ParfunB}({\bm \varphi}_{i-1}, {\bm \alpha}^{(0)}, {\bm \beta}^{(0)}, {\bf  W}^{(0)}; \xi_i)}, \label{equ:ssca_grad2}
\end{align}
\end{subequations}
and $\kappa_i>0$ should be properly chosen to satisfy \eqref{equ:ssca_c2}.
One can see that, given ${\bm \varphi}_{i-1}$ and channel realizations about $\xi_i$, the surrogate function ${\hat h}_i({ {\bm \varphi}}, { {\bm \varphi}_{i-1}}; \xi_i)$ in \eqref{equ:Surrogate_fun} has closed-form expression, and the gradient of $h_r({\bm \varphi})$ is $\frac{1}{r} \sum_{i=1}^r \nabla {\hat g}({\bm \varphi}_{i-1}; \xi_i )$.
\subsubsection{Gradient Projection for Updating ${\bm \varphi}_r$}
The main drawback of the sample average approximation in \eqref{equ:saa} is that, the upadte of ${\bm \varphi}_r$ is related to all ${\hat g}({\bm \varphi}; \xi_i )$ for $i=1,\cdots,r$.
In \cite[Theorem 1]{LiuATSP2018onlineSSCA}, a better recursive approximation is proposed which can be exploited to further simplify the update rule.
%The above stochastic SCA method can be further simplified according to \cite[Theorem 1]{LiuATSP2018onlineSSCA}.
To be specific, $f_{{\rm B},r}({\bm \varphi})$ can be approximated recursively by $h_r$ as follows:
\begin{equation}\label{equ:Recursive_obj}
h_r=(1-\delta_{r}) h_{r-1}+\delta_{r} {\hat g}({\bm \varphi}; \xi_r ),
\end{equation}
where $h_{0}=0$, and we set $\delta_{r}=r^{-0.501}$.
Then, replacing $\frac{1}{r}$ with $\delta_r$, the gradient of $f_{{\rm B},r}({\bm \varphi})$ is approximated by
\begin{equation}\label{equ:Recursive_gradient}
{\bf h}_r=(1-\delta_{r}) {\bf h}_{r-1}+\delta_{r} \nabla {\hat g}({\bm \varphi}; \xi_r ),
\end{equation}
with ${\bf h}_{0}={\bm 0}$. Combining \eqref{equ:Surrogate_fun}, the surrogate function for $h_r$ can be also expressed by a recursive formula as
\begin{equation*}%\label{equ:Surrogate_fun2}
\begin{aligned}[b]
{\bar h}_r({\bm \varphi})&=(1-\delta_{r}) h_{r-1}+ \delta_{r} {\hat g}({\bm \varphi}_{r-1}; \xi_r )\\
&\quad +{\bf h}_r^{\rm T} ({\bm \varphi}-{\bm \varphi}_{r-1})+\frac{\kappa_r}{2}\|{\bm \varphi}-{\bm \varphi}_{r-1}\|^2.
\end{aligned}
\end{equation*}
Finally, we have following update rule for ${\bm \varphi}_r$:
\begin{equation}\label{equ:B_update_varphi}
{\bm \varphi}_r={{\bm \varphi}_{r-1}}-\frac{{\bf h}_r}{\kappa_r},
\end{equation}
which is similar to the gradient projection updating in  \eqref{equ:A_update_varphi}, and thus $\kappa_r$ can be determined by the Armijo rule.

\subsection{Algorithm Development}
Putting all above together, when  a new realization $\xi_r$ is obtained, we first calculate the gradient $\nabla {\hat g}({\bm \varphi}_{r-1}; \xi_r )$ according to \eqref{equ:ssca_grad1} and \eqref{equ:ssca_grad2}.
Then, ${\bf h}_r$ is updated according to \eqref{equ:Recursive_gradient}.
Substituting ${\bf h}_r$ into \eqref{equ:B_update_varphi}, we have ${\bm \varphi}_r$.
%For simplicity, the stop criterion is designed as when the average of  temporary output $f_r=\frac{1}{r} \sum_{i=1}^r h({\bm \varphi}_i, {\bm \varphi}_{i-1}; \xi_i )$ converges.

The proposed  stochastic SCA approach is summarized in Algorithm \ref{alg:P2}.
Denote by $I_{\rm A}$ the iteration number of the Armijo search, by $I_{\rm S}$ the iteration number in Algorithm \ref{alg:ParfumB}.
The complexity of  the proposed algorithm is ${\mathcal O}\left(I_{\rm O}\left(I_{\rm A}I_{\rm S} \left(2 K N M+K M^2\right)+  K^2 N^2\right)\right)$. Hence, compared with Algorithm \ref{alg:P1}, Algorithm \ref{alg:P2} cost $I_{\rm S}$ times complexity with respect to $M$ to obtain a stationary beamforming solution in every outer-loop iteration. However, the complexity with respect to $N$ is still about ${\mathcal O}\left(I_{\rm O}  K^2 N^2\right)$.

\begin{algorithm}[!t]
\caption{Stochastic SCA for the imperfect CSI setup.}
\label{alg:P2}
\begin{algorithmic}[1]
%\REQUIRE
% channel coefficients ${\bf G}$, ${\bf h}_{{\rm d},k}$ and ${\bf h}_{{\rm r},k}$ for all $k$,  and the transmit power constraint $P_{\rm T}$.
\STATE {Initialize ${\bm \varphi}_0$ and set $r=0$.\\
{\bf Repeat}}
\STATE  Set $r=r+1$ and obtain new channel realizations ${\bf h}_{{\rm d},k}(\xi_r)$, ${\bf G}(\xi_r)$, ${{\bf h}_{{\rm r},k}}(\xi_r)$ for all $k$.
\STATE  Calculate the gradient $\nabla {\hat g}({\bm \varphi}_{r-1}; \xi_r )$ by \eqref{equ:ssca_grad1} and \eqref{equ:ssca_grad2}.
\STATE  Update ${\bf h}_r$ by \eqref{equ:Recursive_gradient}.
\STATE  Search $\kappa_r$ in \eqref{equ:B_update_varphi} by  Armijo rule, and then update ${\bm \varphi}_r$.
\\
{\bf Until} The value of $h_r$ in \eqref{equ:Recursive_obj} converges.
\end{algorithmic}
\end{algorithm}

\begin{table*}[!t]
%\scriptsize
\footnotesize
\renewcommand{\arraystretch}{1.3}
\caption{Simulation Parameters}
\label{tablepm}
\centering
\begin{tabular}{c|c}
\hline
Parameters & Values \\
\hline
AP location & ($0$m, $0$m)\\
%\hline
% Small scale fading  $\forall i, k, j$ & ${[{\bf h}_{{\rm d},k}]_i}$, $[{\bf G}]_{ij}$, ${[{\bf h}_{{\rm r},k}]_i}$ $\sim {\cal{CN}}(0,1)$\\
\hline
Path-loss for $\bf G$ and ${{{\bf h}}_{{\rm r},k}}$ (dB)& $35.6 + 22.0 \lg d$\\
\hline
Path-loss for ${{\bf h}}_{{\rm d},k}$ (dB)& $32.6 + 36.7 \lg d$\\
\hline
 Transmission bandwidth  & $180$ kHz\\
\hline
Noise power spectral density & $-170$ dBm/Hz\\
\hline
\end{tabular}
\label{table_sim}
\end{table*}

\textcolor{black}{\section{Numerical Results}\label{simulation}}
\subsection{Simulation Scenario}\label{simulation_link}
In this section, numerical examples are provided to validate the effectiveness of the proposed algorithms.
We consider a RIS-aided femtocell network illustrated in {\figurename~\ref{simulation_scena}}, in which one AP equipped with $4$ antennas, and $4$ single-antenna users ($K=4$) uniformly and randomly distributed in a circle centered at ($200$ m$,30$ m) with radius $10$ m.
The RIS is applied to provide high-quality link between the AP and users, and we assume that the LOS component is contained by the channel between AP and RIS, and channel between RIS and each user.
The system parameters are summarized in Table \ref{table_sim}, which are almost the same as those in \cite{YuenChauIRS}.
In particular, the path-loss is set according to the 3GPP propagation environment \cite[Table B.1.2.1-1]{3GPP}.

\begin{figure}
[!t]
\centering
\includegraphics[width=.9\columnwidth]{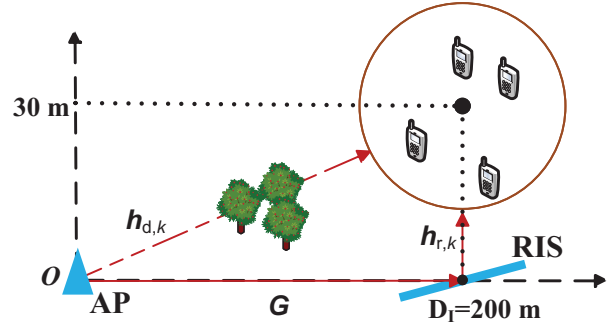}
\caption{The simulated RIS-aided $K$-user MISO communication scenario comprising of one $M$-antenna AP and one $N$-element RIS.}
\label{simulation_scena}
%\vspace{-1.5em}
\end{figure}

We assume the direct link channel ${{\bf h}}_{{\rm d},k}$ follows Rayleigh fading, while the RIS-aided channels follow Rician fading.
Same as \cite{Jinshi2019TVTsurfaceAverage} and \cite{YuanXJ2019LISChannelestimation}, we further assume that the antenna elements form a half-wavelength uniform linear array configuration at the AP and the RIS, and thus the channels $\bf G$ and ${{{\bf h}}_{{\rm r},k}}$ are modeled by
%\begin{align}
%%\begin{equation*}
%{\bf G}&=L_{1} \sum_{i=1}^{\ell_1} \varepsilon_{0,i} {\bf a}_N(\vartheta_i){\bf a}_M(\psi_i)^{\rm H},\\
%{{{\bf h}}_{{\rm r},k}} &=L_{2,k} \sum_{j=1}^{\ell_2} \varepsilon_{k,j} {\bf a}_N(\varsigma_{k,j}),
%%\end{equation*}
%\end{align}
\begin{align}
%\begin{equation*}
{\bf G}&=L_{1} \left(\sqrt{\frac{\varepsilon}{\varepsilon+1}} {\bf a}_N(\vartheta){\bf a}_M(\psi)^{\rm H}+
\sqrt{\frac{1}{\varepsilon+1}} {\bar {\bf G}}\right),\\
{{{\bf h}}_{{\rm r},k}} &=L_{2,k} \left( \sqrt{\frac{\varepsilon}{\varepsilon+1}} {\bf a}_N(\varsigma_{k})+
\sqrt{\frac{1}{\varepsilon+1}} { {{\bar{\bf h}}_{{\rm r},k}}} \right),
%\end{equation*}
\end{align}
where $L_1$ and $L_{2,k}$ denote the corresponding path-losses, $\varepsilon$ is the Rician factor and we set $\varepsilon=10$, ${\bf a}$ is the steering vector, $\vartheta$, $\psi$ and $\varsigma_{k}$ are the angular parameters, and ${\bar {\bf G}}$ and ${{\bar{\bf h}}_{{\rm r},k}}$ denote the NLOS components whose elements are chosen from ${\cal{CN}}(0,1)$.

%, with $\sum_i \sigma_{0,i}^2=1$ and $\sum_j \sigma_{k,j}^2=1$, $\ell_1$ and $\ell_2$ are the number of propagation paths, and we assume $\ell_1=6$ and $\ell_2=35$  \cite{YuanXJ2019LISChannelestimation}.

Based on above assumption, only the small-scale fading variables ${{\bf h}}_{{\rm d},k}$, ${\bar {\bf G}}$, and ${{\bar{\bf h}}_{{\rm r},k}}$ need to be estimated in every frame.
Denote $x$ as one element in above variables, and $\hat x$ is the corresponding estimate value.
We assume that the estimate error $x-\hat x$ follows zero mean complex Gaussian distribution, and all these elements have the same normalized MSE:
\begin{equation*}
\varrho=\frac{{\mathbb E} \left[\left|x-\hat x\right|^2\right]}{{\mathbb E} \left[\left|\hat x\right|^2\right]}
.
\end{equation*}

To better understand the channel conditions of the direct link and the RIS-aided link, we provide a simple example here.
Consider a reference point at ($200$m, $30$m).
According to Table \ref{table_sim}, the direct-link path-loss  is about $117.23$ dB, meanwhile, the path-loss of channel $\bf G$ and channel ${{\bf h}}_{{\rm r}}$ are $86.22$ dB and $68.10$ dB, respectively, so the path-loss of the RIS-aided link ($N=1$) is $154.32$ dB, which is much larger than that of the direct link (about $37$ dB).
Therefore, the direct link cannot be ignored, and extremely large $N$ is required to achieve performance gain, if the surface phase vector $\bm \theta$ is not properly designed.
%This is reasonable, since based on the scatter radio principle \cite{Griffin2009LinkBudget}, most energy of the re-scatter signal from the RIS cannot arrive the user's receiver without proper designed surface phase vector.
In the next, we will show that, by utilizing the proposed joint optimization algorithms, significant performance gain can be achieved.

We evaluate the performance of the proposed algorithms with the following $3$ baselines:
\begin{itemize}
\item {\bf Baseline 1} (Without RIS): Let $N=0$, and then ${\mathcal{P}}{\text{(A)}}$ is solved by  the WMMSE.
\item {\bf Baseline 2} (Random Phase): $\bm \theta$ is initialized by random value, and then $\bf W$ is optimized by WMMSE.
\item {\bf Baseline 3} (Upper Bound): The KKT conditions are necessary conditions for a solution to be optimal. Thus one may run Algorithm \ref{alg:P1} sufficient times (e.g., $100$ times) with random initializations, and then the maximum output might approximate the optimal solution well.
\end{itemize}

\subsection{Weighted Sum Rate Analyses}
In this subsection, we assume that the RIS is deployed at ($200$m, $0$m), and the users' locations are fixed once randomly generated.\footnote{In the simulation, the user locations are ($205.65$m, $34.48$m), ($193.47$m, $30.24$m), ($198.30$m, $22.40$m), and ($207.00$m, $24.28$m).}
Then, for fairness comparison, the weights are first chosen inversely proportional to the direct-link path-loss, and then normalized by $\sum \omega_k=1$. %\footnote{Thus the weights are irrelevant to $N$.}
All the simulation curves have been averaged over $10^3$ independent realizations of channel small scale fading.

\begin{figure}
[!ht]
  \centering
  \subfigure[$P_{\rm T}$ vs WSR]{
    \label{wsr_vs_PT:a} %% label for first subfigure
    \includegraphics[width=1\columnwidth]{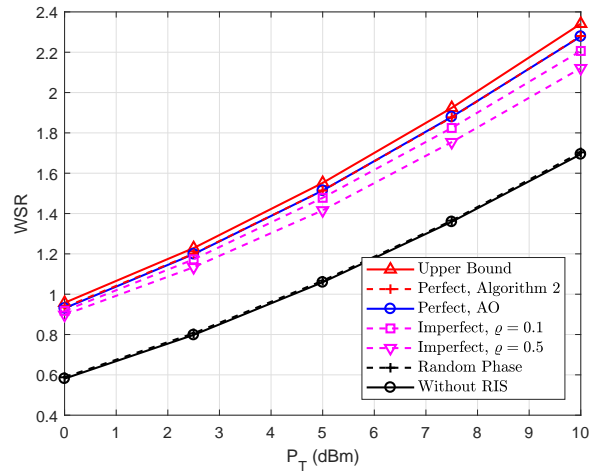}}
  %\hspace{1in}
  \subfigure[Convergence behavior when $P_{\rm T}=0$ dBm]{
    \label{wsr_vs_PT:b} %% label for second subfigure
    \includegraphics[width=1\columnwidth]{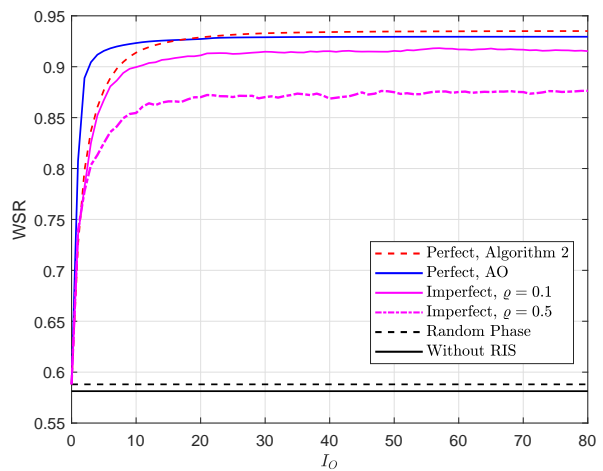}}
  \caption{The WSR versus transmit power when $N=100$.}
  \label{wsr_vs_PT} %% label for entire figure
\end{figure}

{\figurename~\ref{wsr_vs_PT:a}} illustrates the WSR of different schemes with respect to the transmit power $P_{\rm T}$ when $N=100$.
%The two baselines in black curves are the WSR obtained by the system without RIS and the RIS-aided system with random phase vector, respectively.
%Then, all the other curves are initialized by the random-phase scheme.
It is seen that, if the phase vector is not optimized, the performance gain by deploying RIS is negligible as expected.
However, the joint beamforming and phase optimization schemes may achieve about $4$ dB gain.
In addition, in perfect CSI setup, the proposed Algorithm \ref{alg:P1} and the alternating optimization approach have almost the same performance. We conjecture that this is due to both algorithms are initialized by the same point, and then they both converge to the same stationary solution with high probability.
In imperfect CSI setup, one can see that, the proposed algorithm may still achieve about $3$ dB gain when $\varrho\leq 0.5$.
Besides, the performance loss increases as $P_{\rm T}$ increases.
%However, when $\varrho= 0.5$, the performance gain decreases remarkably.

Next, in {\figurename~\ref{wsr_vs_PT:b}}, we fix the transmit power $P_{\rm T}=0$ dBm and show the convergence behaviors of all the proposed algorithms. %by illustrating the output value of the objective function in each outer-loop step.
In perfect CSI setup, the convergence speed of the proposed algorithm is slightly slower than the alternating optimization approach, which the performance is sightly better. In addition, as we have analyzed in Section \ref{Sec:development_A}, in each iteration,  the proposed algorithm has much lower  complexity.
In imperfect CSI setup, one can see that, as $\varrho$ increases, the channel becomes more uncertain, and the proposed algorithm needs more steps to get converged.

%Interestingly, the proposed Algorithm \ref{alg:P2} for imperfect CSI setup with $\varrho=0.01$ has the quickest convergence speed.
%This is owing to that, in each outer-loop iteration, Algorithm \ref{alg:P2} takes an additional inner-loop search to find a better beamforming solution.
%However, when $\varrho$ increases, the convergence speed becomes slower, as the difference between two arrived channel realizations becomes larger.

%\footnote{The proposed algorithm for the imperfect CSI setup could be done off-line by estimating $\varrho$ from historical observations which is also a long-term variable determined by the specific channel estimation technique and the dynamic of the channel.}

{\figurename~\ref{N_vs_PT}} compares the WSR with the size $N$ of  RIS, while the transmit power of AP is fixed to $P_{\rm T}=5$ dBm.
The random-phase scheme still has only a small gain, meanwhile all the schemes optimized $\bm \theta$ achieve remarkable performance gain as $N$ increases.
Besides, we observe that, the performance of the proposed algorithm at $N=200$ is similar to that at $P_{\rm T}=8$ dBm in {\figurename~\ref{wsr_vs_PT:a}}.
This observation implies that, different from \cite[Proposition 2]{zhangruiIRS}, the RIS phase design could not achieve the ``squared gain'' here, since the aperture gain of the RIS is relatively small.

\begin{figure}
[!t]
\centering
\includegraphics[width=1\columnwidth]{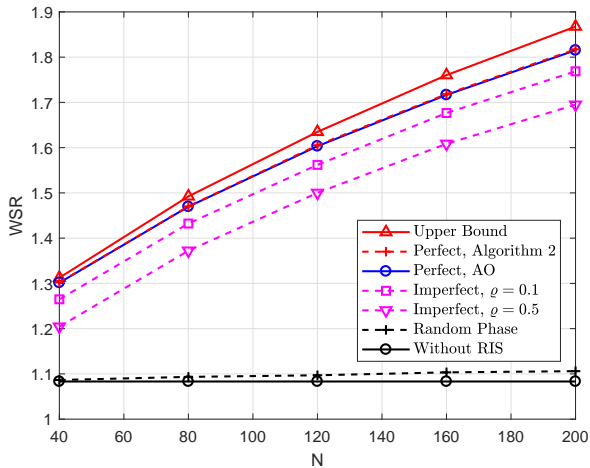}
\caption{WSR versus $N$, when $P_{\rm T}=5$ dBm.}
\label{N_vs_PT}
%\vspace{-1.5em}
\end{figure}

\begin{figure}
[!t]
\centering
\includegraphics[width=1\columnwidth]{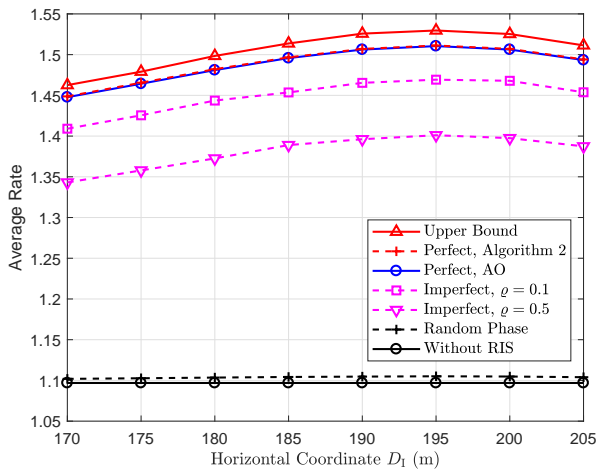}
\caption{The average rate versus the horizontal coordinate  of RIS, when $P_{\rm T}=5$ dBm and $N=100$.}
\label{location_vs_R}
%\vspace{-1.5em}
\end{figure}

\subsection{RIS Deployment and User Locations}
In this subsection,  we discuss on the impact of the RIS deployment locations and users' locations, and the horizontal coordinate of RIS is denoted by $D_{\text I}$.
The weights are set to be equal to $1/K$, so the objective function becomes average rate.
We generate $100$ snapshots for randomly located users.
Then, for each snapshot, we further generate $100$ channel realizations with independent small-scale fading.

{\figurename~\ref{location_vs_R}} illustrates the average rate of users when $P_{\rm T}=5$ dBm and $N=100$, while moving the RIS from ($170$m, $0$m) to ($205$m, $0$m).
It is seen that, when $D_{\text I}$ increases from $200$ m to $205$ m, the average rate decreases, since the path-losses of channel $\bf G$ and channel ${{\bf h}}_{{\rm r}}$ both increase.
However, when decreasing $D_{\text I}$ from $200$ m to $170$ m, the average rate first increases, and then decreases, while the optimal location is $D_{\text I}=195$ m.
This is because, the path-loss of the RIS-aided link is the product of the  path-losses of $\bf G$ and ${{\bf h}}_{{\rm r}}$.
Therefore, although the summation of the transmission distance decreases, the propagation condition might not necessarily become better.

%the performance gain of the RIS-aided system increases when the RIS is deployed closer to the AP, since the path-loss from AP to RIS decreases while the change about the path-loss from RIS to users is small.
%However, it should be noted that, the propagation condition between RIS to users may get as worse as the direct link, if we continue moving the RIS closer to the AP.

Finally, {\figurename~\ref{CDF_location}} plots the \emph{cumulative distribution function} (CDF) of the average rate over different snapshots by deploying RIS at ($195$m, $0$m).
It is seen that, the performance gains of all the proposed schemes are stable over the CDF curves, and also keep consistent with their counterparts in {\figurename~\ref{location_vs_R}}.
Therefore, we conclude that, with high probability, the performance of the proposed algorithms will be good irrespective of user locations.

\begin{figure}
[!t]
\centering
\includegraphics[width=1\columnwidth]{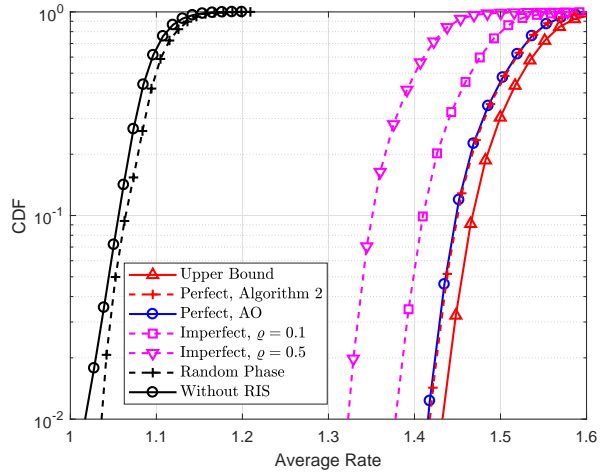}
\caption{CDF curves for random user locations.}
\label{CDF_location}
%\vspace{-1.5em}
\end{figure}

\section{Conclusion}\label{conclusion}
In this paper, we investigate the  RIS-aided multiuser downlink MISO system.
Specifically, a joint transmit beamforming design and RIS phase optimization problem is formulated to maximize the WSR under the AP transmit power constraint.
The perfect CSI setup is firstly addressed, and a low-complexity algorithm is designed to carry out stationary solution for the joint design problem by utilizing the recently proposed  FP technique.
The proposed algorithm is then leveraged to the imperfect CSI setup, and the average WSR is maximized by resorting to the stochastic SCA technique.
Extensive simulation results demonstrated that the proposed joint design schemes achieve significant performance gain compared to the benchmarks by deploying a RIS with $100$ passive elements.
In addition, it is also shown that the performance degradation of the proposed algorithm is very small, when the channel estimation uncertainty is smaller than $10\%$.

\bibliographystyle{IEEEtran}%By using IEEEtrans, the number can be displayed.
%\bibliography{ref_AmBC}
\bibliography{IEEEabrv,mybib_draft2}

\end{document}